\documentclass[11pt]{article}
\usepackage{setspace}
\usepackage{graphicx}
\usepackage[space]{grffile}
\usepackage{amsmath}
\usepackage{amsthm}
\usepackage[sort,comma,authoryear]{natbib}
\usepackage{mathtools}
\usepackage{bm}
\usepackage{amssymb}
\usepackage{threeparttable}
\DeclarePairedDelimiter{\abs}{\lvert}{\rvert}
\usepackage{booktabs}
\usepackage{array}
\usepackage[english]{babel}
\usepackage{longtable}
\usepackage{subfig}
\usepackage{url}
\usepackage{ragged2e}
\usepackage{tabularx}
\usepackage{epstopdf}
\usepackage[toc,title,page]{appendix}
\usepackage{longtable}
\usepackage{threeparttable}
\usepackage{rotating}

\theoremstyle{definition}

\theoremstyle{remark}

\usepackage{tabularx}
\usepackage{changepage}
\usepackage{verbatim}
\usepackage[a4paper, total={6in, 8in}]{geometry}
\usepackage{setspace}
\usepackage{booktabs}
\usepackage{geometry}
\usepackage{multirow}
\numberwithin{equation}{section}
\usepackage{geometry}
\newcommand{\tabincell}[2]{\begin{tabular}{@{}#1@{}}#2\end{tabular}}

\usepackage{hyperref}
\usepackage{natbib}
\hypersetup{
	colorlinks,
	linkcolor={blue},
	citecolor={blue},
	urlcolor={blue}
}

\usepackage{algorithm}
\usepackage{algorithmicx}
\usepackage{algpseudocode}

\usepackage{indentfirst}
\begin{document}
\date{}

\title{Analyzing covariate clustering effects in healthcare cost subgroups: insights and applications for prediction}


\author{
	Zhengxiao Li \footnote{School of Insurance and Economics, University of International Business and Economics, Beijing, China.}
	\quad Yifan Huang \footnote{corresponding author: School of Insurance and Economics, University of International Business and Economics, Beijing, China. Email: huangyf2217@163.com.}
	\quad Yang Cao
	\footnote{School of Insurance and Economics, University of International Business and Economics, Beijing, China. }
}

\maketitle

\begin{abstract}
Healthcare cost prediction is a challenging task due to the high-dimensionality and high correlation among covariates. Additionally, the skewed, heavy-tailed, and often multi-modal nature of cost data can complicate matters further due to unobserved heterogeneity. In this study, we propose a novel framework for finite mixture regression models that incorporates covariate clustering methods to better account for the effects of clustered covariates on subgroups of the outcome, which enables a more accurate characterization of the complex distribution of the data.
The proposed framework can be formulated as a convex optimization problem with an additional penalty term based on the prior similarity of the covariates. To efficiently solve this optimization problem, a specialized EM-ADMM algorithm is proposed that integrates the alternating direction multiplicative method (ADMM) into the iterative process of the expectation-maximizing (EM) algorithm. The convergence of the algorithm and the efficiency of the covariate clustering method are verified using simulation data, and the superiority of the approach over traditional regression techniques is demonstrated using two real Chinese medical expenditure datasets. Our empirical results provide valuable insights into the 
complex network graph of the covariates
and can inform business practices, such as the design and pricing of medical insurance products.
	\\
	{\bf{Keywords:}} covariate clustering; finite mixture regression models; EM-ADMM algorithm; healthcare costs prediction.
	\\
	\\
	{\bf{Article History:}}  \today
\end{abstract}

\newpage

\section{Introduction}\label{section-intro}
The prediction of healthcare costs for individuals is a crucial risk management concern for healthcare systems and health insurers in determining appropriate premiums \citep{fellingham2015bayesian, xie2016analyzing}. While regression methods have been well-developed for predictive modelling of healthcare costs from individual clinical data in the economic and biomedical statistics literature \citep{bertsimas2008algorithmic, neelon2015spatiotemporal, cao2012forecasting, duncan2016testing}, high-dimensional regression models are often necessary due to the large number of risk factors in healthcare datasets, some of which are highly correlated and grouped into sets (or clusters) known to the data analyst (e.g. demographic information such as age, gender, income; health status such as smoking, drinking, surgery history; and clinical measurements such as blood pressure, blood glucose, resting heart rate). Despite successful covariate selection, the covariates need further clustering to increase model accuracy and interpretability \citep{andrade2021convex}. Another challenge is that covariates respond differently to the response variable due to unobserved heterogeneity and latent factors, and common distributions such as Gamma, Weibull, and Lognormal may not fully capture the heavy-tailedness and multi-modality of individual healthcare costs data \citep{mihaylova2011review}, which also suggests the need to
identify the subgroups of the outcome data.
To address these challenges, this paper proposes a statistical regression model for healthcare cost prediction, with the aim of precise covariate selection and covariate clustering for heterogeneous subgroups of individuals.

Given the heterogeneity which characterizes individual clinical data, many approaches have been proposed in the statistical literature to identify unobserved subgroups from a heterogeneous population \citep{ma2017concave}.
One such approach is to treat the outcome data as a mixture of several subgroups with their own parameters and to fit a finite mixture regression model (FMR). Previous studies have explored FMR models with no covariates \citep{ickowicz2017modelling, kurz2019identifying}, as well as those with low-dimensional covariates \citep{shen2015inference}. While mixture models can easily incorporate covariate effects due to their theoretical framework, they are limited when covariates are highly correlated in high-dimensional problems. To address this issue, regularization approaches have been developed in FMR models \citep{khalili2007variable, khalili2013regularization}, with the $\ell_p$ regularization term being a particularly popular choice. More recently, a high-dimensional mixture Gamma regression model was proposed to fit heavy-tailed insurance claim data with neural networks, allowing for covariate selection and detection of non-linear regression patterns \citep{devriendt2021sparse}.
This indicates a mixture Gamma distribution is preferred to model the right-skewed, heavy-tailed and multi-modal insurance claim data \citep{yin2016efficient, miljkovic2016modeling}.
However, these methods cannot reveal the structure of the relationship between highly correlated covariates, although they can reduce the dimensionality of the data.
Therefore, further clustering methods for highly correlated covariates are still needed to improve the interpretability of the model.

One commonly used approach to reducing the dimensionality of covariates is unsupervised clustering.
For instance, \cite{tsai2021incorporating} applied four clustering approaches (i.e., Ward’s hierarchical clustering, divisive hierarchical clustering, K-means clustering, and Gaussian mixture model clustering) to improve forecasting performance of mortality models in insurance aspect.
The drawback of these methods is that it is generally performed independently of regression, overlooking the response-covariate relationships, leading to undesirable clustering results.
A approach for identifying clusters among the covariates and simultaneously
estimating the regression coefficients has been proposed in several studies, such as \cite{witten2014cluster}, \cite{sui2018convex}, \cite{hallac2015network} and \cite{andrade2021convex}.
The key idea of \cite{andrade2021convex} is that while taking into account the response information is to formulate a convex
classification problem with a pair-wise group lasso penalty, which selectively shrinks the
coefficients of such covariates that are highly correlated but not related to the origin.
This modelling framework can be also referred to
a network lasso \citep{hallac2015network} or generalized lasso problem \citep{zhu2017augmented}.
However, the existing methods have limitations in identifying and interpreting different clustering results from various subgroups of healthcare costs data.

In this paper, we present a new statistical framework for healthcare cost modeling that identifies the contribution of clustered covariates to the subgroups of the outcome, with clustering results varying across different components of the FMR model.
The proposed approach builds on the work of \cite{andrade2021convex}, which clusters the regression coefficients of covariates using pairwise fused penalty.
Specifically, the framework starts with an FMR model based on the exponential distribution family, where the outcome is treated as a mixture of different subgroups. Each subgroup is assumed to be an underlying distribution with its own set of parameters.
The mean parameter of each underlying distribution is formulated by linearly combining the covariates using a log-link function.
To obtain the covariate clustering results during mixture regression, prior similarity information about the covariates is used, and two convex penalty terms are added to the log-likelihood function. These penalty terms account for the fit deviation of observed samples (the $\ell_2$ norm of the coefficients) and the distance of similarity between covariates (the $\ell_1$ norm of the difference of pair-wise coefficients) in different subgroups.
The proposed model adaptively explores the pair-wise structure of a large number of highly correlated covariates by shrinking the group-specific coefficients in different mixture components. The approach can cluster the covariates information when subgroup regression is performed simultaneously,
which is an improvement over traditional methods that separate the two steps and model them disjointly. The proposed framework can also be applied to one-dimensional or non-mixture models when there is weak heterogeneity within the observed sample. Notably, the proposed model can be viewed as an extension of the linear regression models with convex clustering method proposed by \cite{cheng2022robust}.

Our second contribution is employing a specialized EM-ADMM algorithm for optimizing the complicated objective function.
In this algorithm, we perform an alternating direction method of multipliers (ADMM) at the M-step of the Expectation-Maximization (EM) algorithm, instead of using the Newton-Raphson optimization procedure to estimate the unknown parameters. The ADMM algorithm has good convergence properties for convex objective functions with the $\ell_p \ (p\ge1)$ regularization terms, as demonstrated by \cite{boyd2011distributed} and \cite{chi2015splitting}.
Furthermore, the $\ell_1$ regularization can shrink some pairwise differences of the parameter estimates to zero. To obtain satisfactory covariate clustering results for each component in the mixture model, we choose the penalty imposed on the regression coefficients within each component of the FMR model to be proportional to the mixture weight, following the work of \cite{khalili2007variable}.
Our simulation study shows that the proposed algorithm provides a parallel computational strategy for the parameter estimation of each component in the mixture model, resulting in a fast convergence rate and high accuracy.

We present two practical healthcare costs datasets to demonstrate the effectiveness of our proposed model.
The first dataset pertains to a health insurance product developed for chronic diabetes, with the primary goal being the prediction of individual medical costs associated with diabetes based on personal patient information.
This dataset is unique in that it contains a questionnaire consisting of 111 questions, many of which are similar and can be grouped into specific sets of highly correlated covariates measuring self-perception of diabetes for each patient.
Given that medical utilization costs for diabetes patients are strongly correlated with comorbidities and complications, patients with a better understanding of diabetes are more likely to receive timely treatment, including regular glucose-lowering medication or insulin injections, thus reducing the probability of complications and related costs.
Our motivation to apply the covariate clustering method is to uncover the relationship structures within the questionnaire data, which can help improve the model's predictive performance.
The second dataset is short-term medical insurance claims data provided by a Chinese insurance company, where the ICD codes can be clustered into specific groups for different subgroups during mixture regression.
Our empirical results demonstrate that our proposed model is suitable for heavy-tailed and multi-modal medical costs data, and that the combination of regression and covariate clustering can significantly improve model accuracy and interpretability for predicting future insurance claims.
This can be particularly useful in guiding business practices such as designing and pricing medical insurance products.

The rest of this paper is structured as follows:
in Section \ref{section-FMR-clustering}, we first define the framework of FMR models with covariate clustering, including the symbols and main objective function.
An EM-ADMM algorithm for solving the optimization problem as well as some computation details are then shown in Section \ref{section-algorithm}.
Section \ref{section-simulation} verifies the performance of proposed model through simulation studies, while in Section \ref{section-empirical} empirical analyses of two real datasets are illustrated.
Conclusions are given in Section \ref{section-conclusion}.
To make the framework and algorithm more accessible to practitioners, 
we provide the R programming codes for implementation at \url{https://github.com/huangyf2217/FMR-covariates-clustering}.

\section{Finite mixture of regression model with covariate clustering}\label{section-FMR-clustering}
\subsection{Finite mixture of regression models}
Finite mixture models provide a tool for analyzing data suspected of coming from a heterogeneous population.
The positive continuous response variable of interest (e.g., healthcare costs) is consistently denoted as $Y$, and $\bm{x} = (x_1,x_2, ...,x_p)$ is the $p$-dimensional covariate vector related to the risk.

Generally, the unobservable latent factors in heterogeneous response variable can be distinguished by the mixture of underlying distributions.
We first assume that the random variable $Y$ comes from a parametric exponential family $\mathbb{F} = \left\{ f(y; \mu,\phi); (\mu, \phi) \in \mu \times \left(0, +\infty \right) \right\}$, where $\mu$ is the specific mean parameter and $\phi$ is the dispersion parameter.
In generalized linear models (GLMs), the mean parameter can be expressed as a function related to the covariates, specifically, $\mu:=\mu(\bm{x})$.
Then, the conditional density function of $Y$ given $\bm{x}$ has the form, extending the above structure to an FMR model:

\begin{equation}
\label{eq-FMR-density}
f_{\text{mix}} \left(y \left| \bm{x}, \bm{\omega}, \bm{\beta}_0, \mathbf{B}, \bm{\phi} \right. \right)
=\sum_{h=1}^H {\omega_h} {f\left(y \left| \mu_h(\bm{x}), \phi_h \right. \right)},
\end{equation}
where $\mu_h(\bm{x}) = g^{-1}(\beta_{0,h} + \bm{x}^T \bm{\beta}_h)$ with a link function $g(\cdot)$.
$\bm{\beta}_0 = \left(\beta_{0,1},\beta_{0,2},\dots,\beta_{0,H} \right)^T \in \mathbb{R}^{1\times H}$ and
$\mathbf{B} = (\bm{\beta}_1,  \bm{\beta}_2,  \dots , \bm{\beta}_H)^T \in \mathbb{R}^{p\times H}$  contain the intercept terms and regression coefficients respectively, where
$\bm{\beta}_h = (\beta_{1,h}, \beta_{2,h}, \dots, \beta_{p,h})$ is the vector of regression coefficients in the $h$ component.
$\bm{\phi} = \left(\phi_1,\phi_2,\dots,\phi_H \right)^T\in \mathbb{R}^{1\times H}$ represents the dispersion parameters of all $H$ components.
The component weight $\bm{\omega} = \left(\omega_1,\omega_2,\dots,\omega_H \right)^T$ satisfies the constraints $\omega_h > 0$ and $\sum_{h=1}^{H}\omega_{h}=1$, so the mixture density function is actually the weighted sum of all components' densities.
The form of $f(\cdot)$ can vary according to the response variable. Specifically, a Gamma distribution is often chosen as the underlying distribution in FMR for fitting the positive continuous data in insurance applications.

Given a sample of observations $(\bm{x}_1,y_1),(\bm{x}_2,y_2),...,(\bm{x}_n,y_n)$,
the (penalized) log-likelihood function for the FMR model is as follows:
\begin{equation}
\label{eq-FMR-penalized-loglike}
\breve{\ell}_n(\bm{\Psi}) = \ell_n(\bm{\Psi}) - p_n(\bm{\Psi}),
\end{equation}
and
\begin{equation}
\label{eq-FMR-loglike}
\ell_n(\bm{\Psi})=\sum_{i=1}^n\log f_{\text{mix}} \left(y_i \left| \bm{x}_i, \bm{\omega}, \bm{\beta}_0, \mathbf{B}, \bm{\phi} \right. \right)=\sum_{i=1}^n\log \sum_{h=1}^H {\omega_h}{f\left(y_i \left| \mu_h(\bm{x}_i), \phi_h \right. \right)},
\end{equation}
where
$\bm{\Psi} := \left(\bm{\omega}, \bm{\beta}_0, \bm{\beta}, \bm{\phi} \right)$,
$x_{i}=(x_{1,p},...,x_{i,p})$ is the $p$-dimensional covariates vector,
and
the regularization term is defined as $p_n(\bm{\Psi}) = \gamma \sum_{h=1}^H {\omega_h}{\left\| \bm{\beta}_h \right\|_{2}^{2}}$ with $\gamma$ being a non-negative regularization parameter.
This is in line with the common practice of linking the penalty term to sample size, and improves the model's ability to identify important covariates, given that the weight $\omega_h$ reflects to some extent the size of the $h$th subpopulation \citep{khalili2007variable}.
The weighted $\ell_2$ norm is able to compress the insignificant parameters in each component to avoid the model overfitting.
In addition to quantifying the probability that the observation belongs to each component, the regularized FMR models are able to describe how the covariates are related to the response variable under the influence of various latent factors.
The number of subgroups $H$ and the corresponding underlying distributions are usually pre-specified, which is usually not too large in healthcare costs modelling.

\subsection{The framework of convex covariate clustering}
For analysis of the covariate clustering effects on different components in the FMR model,  we can identify the same (or similar) values of corresponding regression coefficients in each component, thus grouping the ``similar" covariates with similar semantics or values.

Suppose a similarity matrix $\mathbf{S}=\{s_{j,k}\}_{p \times p}$ is given a priori, where the element $s_{j,k}$ defines the similarity measure between the covariate $j$ and $k$ that are known to the experts.
Generally, $s_{j,k} = s_{k,j}$ is held.
In the graphical interpretation of clustering,
the similarity matrix can be regarded as an undirected weighted graph, in which each covariate corresponds to a node in a graph, and an
edge connects nodes $j$ and $k$ whenever $s_{j,k} > 0$.

Following the idea of \cite{andrade2021convex}, we
introduce an additional similarity regularization term in \eqref{eq-FMR-penalized-loglike}, where the penalized objective function is defined as
\begin{equation}
\label{eq-FMR-final-objective}
\widetilde{\ell}_n(\bm{\Psi}) = \breve{\ell}_n(\bm{\Psi}) - \frac{v}{2} \sum_{h=1}^{H}
\sum_{j,k \in I} \omega_{h}  s_{j,k} \abs{{\beta}_{j,h} - {\beta}_{k,h}},
\end{equation}
where $v$ is a non-negative tuning constant, $I = \left\{1,2,...,p\right\}$ is the index set of covariates, $\abs{{\beta}_{j,h} - {\beta}_{k,h}}$ measures the Euclidean distance between any pair of two covariates $j$ and $k$ that both have effects on the $h$th component of the outcome.
The similarity penalty is weighted by the value of $s_{j,k}$.
We can cluster two covariates $j$ and $k$,
such that $\abs{\hat{\beta}_{j,h} - \hat{\beta}_{k,h}}$ is arbitrarily
small when $v$ is sufficiently large, where $\hat{\bm{\beta}_h}=(\hat{\beta}_{1,h},..., \hat{\beta}_{p,h})$ is the maximization of \eqref{eq-FMR-final-objective}.
Note that the two regularization terms imposed on the regression coefficients both are set to be proportional to the component weight $\omega_h$ in \eqref{eq-FMR-final-objective}, resulting in the component-specific clustering result of the covariates.
By simply setting $H=1$ in the objective function above, the proposed framework can also be easily regarded as the traditional GLMs with covariate clustering method.

\section{Parameter estimation method}\label{section-algorithm}
\subsection{The optimization process of EM-ADMM algorithm}
The two regularization terms in the penalized objective function ensure the identifiability of our model. When $\gamma$ and $v$ are greater than zero, the optimization problem \eqref{eq-FMR-final-objective} is strongly convex, thus we are guaranteed to find the global optima by using appropriate algorithms.

For simple FMR models, the Expectation-Maximization (EM) algorithm provides a convenient tool to solve the problem of the subgroup analysis based on the maximum likelihood estimation method \citep{dempster1977maximum}.
This algorithm is generally divided into two steps: First, computing the conditional expectation of the complete log-likelihood function for a given sample data belonging to the specific latent subgroup
(i.e. the E-step);
 second, maximizing the conditional expectation of complete log-likelihood and iteratively updating the estimates of $\bm{\Psi}$ (i.e. the M-step).
However, the commonly used Newton-type method is inadequate for solving the optimization problem with multiple pairwise penalty terms, whereas the difficulty of M-step optimization depends on the form of the expected complete log-likelihood.

In this section, we discuss a numerical optimization method that uses the EM algorithm to estimate the parameters in the FMR model, but incorporates the Alternating Direction Method of Multipliers (ADMM) algorithm in the M-step to solve the maximization problem.
We call it EM-ADMM algorithm.
Consistent with the EM algorithm, we first introduce the latent variables
$u_{i,h}$
into the objective function \eqref{eq-FMR-final-objective},
where
$u_{i,h}$
indicates the class of the $i$th observation, and a certain $u_{i,h}=1$ if and only if $y_i$ comes from the $h$th mixture component. Then the penalized complete log-likelihood function based on the latent variables takes the following form:
\begin{equation}
\begin{aligned}
\label{eq-EM-objective}
\widetilde{\ell}_n^c\left(\bm{\Psi} \right) &=
\sum_{i=1}^n \sum_{h=1}^H u_{i,h} \left[\log \omega_{h} + \log{f\left(y_i \left| \mu_h(\bm{x}_i), \phi_h \right. \right)} \right] \\
&\quad - \gamma \sum_{h=1}^H {\omega_h}{\left\| \bm{\beta}_h \right\|_{2}^{2}}
- \frac{v}{2} \sum_{h=1}^{H} \sum_{j,k \in I} \omega_{h} s_{j,k} \abs{{\beta}_{j,h} - {\beta}_{k,h}}.
\end{aligned}
\end{equation}
%

Our proposed algorithm consists of two layers of loops to maximize $\widetilde{\ell}_n^c\left(\bm{\Psi} \right)$. The outer layer is a classic EM iteration that repeatedly implements E-steps and M-steps.
Assume that $m=1,2,\dots,M$ represents the the number of EM iterations.
The EM algorithm at the $(m+1)$th iteration is as follows:

(1) \textbf{E-step}. The E step computes the conditional expectation
of the function $\widetilde{\ell}_n^c\left(\bm{\Psi} \right)$ with respect to $u_{i,h}$, given the data $(\bm{x}_i, y_i)$ and assuming that the current estimate $\bm{\Psi}^{(m)}$ gives the true parameters of the model.
The conditional expectation  (Q-function) is
\begin{equation}
\begin{aligned}
\label{eq-EM-Q-function}
Q(\bm{\Psi} ; \bm{\Psi}^{(m)}) &=
\sum_{i=1}^n \sum_{h=1}^H \pi_{i,h}^{(m)} \log{\omega_h}
+ \sum_{i=1}^n \sum_{h=1}^H \pi_{i,h}^{(m)} \log{f\left(y_i | \mu_h(\bm{x}_i), \phi_h \right)} \\
&- \gamma \sum_{h=1}^H {\omega_h}{\left\| \bm{\beta}_h \right\|_{2}^{2}}
- \frac{v}{2} \sum_{h=1}^{H} \sum_{j,k \in I} \omega_{h} s_{j,k} \abs{{\beta}_{j,h} - {\beta}_{k,h}},
\end{aligned}
\end{equation}
where $\pi_{i,h}^{(m)}$ are the expectations of the unobserved $u_{i,h}$ with the following values:
\begin{equation}
\label{eq-EM-pi}
\pi_{i,h}^{(m)}=\frac{w_h^{(m)} f\left(y_i | \mu_h^{(m)}(\bm{x}_i), \phi_h^{(m)} \right)} {\sum\limits_{h^{'}=1}^{H} w_{h^{'}}^{(m)} f\left(y_i | \mu_{h^{'}}^{(m)}(\bm{x}_i), \phi_{h^{'}}^{(m)} \right)},
\end{equation}
where
$
\label{eq-EM-pi-bx}
\mu_h^{(m)}(\bm{x}_i) = g^{-1}(\beta_{0,h}^{(m)} + \bm{x}_i \bm{\beta}_h^{(m)}).
$

(2) \textbf{M-step}.
The M step at the $(m+1)$th iteration maximizes the function $Q(\bm{\Psi} ; \bm{\Psi}^{(m)}) $ with respect to $\bm{\Psi}$.
We use a sequential method to obtain the parameters' estimates.
First, the component weights $\omega_h$ are updated by:
\begin{equation}
\label{eq-EM-weight}
\omega_{h}^{(m+1)} = \frac{1}{n} \sum_{i=1}^n \pi_{i,h}^{(m)}, \quad h=1,2,\dots,H,
\end{equation}
which maximize the leading term of $Q(\bm{\Psi} ; \bm{\Psi}^{(m)}) $ in \eqref{eq-EM-Q-function}. This is similar to the scheme of \cite{khalili2007variable} and it works well in our simulations.



Next, we fix $\omega_h = \omega_h^{(m+1)}$ and maximize $Q(\bm{\Psi} ; \bm{\Psi}^{(m)})$ with respect to the other parameters in $\bm{\Psi}$.
Since the pairwise fusion penalty involves a large number of parameters and is not differentiable with respect to $\beta_{j,h}$, it is difficult to apply traditional Newton algorithms to realize the M-step optimization. We further consider re-parameterizing the above optimization objective function and estimating the parameters within the framework of the scaled ADMM algorithm\footnote{The details for the scaled ADMM can be found in Appendices \ref{section-appA}-\ref{section-appC}.}.
The inner loops are required in each EM iteration to maximize the Q-function when using the scaled ADMM.
We start from the initial values $\beta_{0,h}^{(m,0)}, \bm{\beta}^{(m,0)}_{h}, \phi^{(m,0)}_{h}$ and update the parameters cyclically for a total of $T$ steps in the $m$th EM iteration.
Let $\bm{z}_{h}=\{z_{j,k,h}\}$ be a set of auxiliary variables and $\bm{r}_{h}=\{r_{j,k,h}\}$  be a set of scaled dual variables for the $h$th component, 
the scaled ADMM algorithm consists of the following steps:
\begin{itemize}
	\item Step 1: Initialize $z_{j,k, h}^{(m,0)} = \beta_{j,h}^{(m,0)}$ and $r_{j,k,h}^{(m,0)} = 0$.
	
	\item Step 2: At the $(t+1)$th iteration of the scaled ADMM, $\beta_{0,h}^{(m,0)}, \bm{\beta}^{(m,0)}_{h}, \phi^{(m,0)}_{h}$ are updated by:
	\begin{align}
	\label{eq-ADMM-beta}
	\mathrm{argmin}_{\beta_{0,h}, \bm{\beta}_{h}, \phi_{h}} \quad g(\beta_{0,h}, \bm{\beta}_{h}, \phi_{h}) \triangleq
	&- \sum_{i=1}^n \pi_{i,h}^{(m)} \log{f\left(y_i | \mu_{h}(\bm{x}_i), \phi_{h} \right)}
	+ \omega_{h} \gamma {\left\| \bm{\beta}_{h} \right\|_{2}^{2}} \nonumber \\
	&+ \frac{\rho}{2} \sum_{j=1}^p \sum_{k=1}^p \left(z_{j,k,h}^{(m,t)} - \beta_{j,h} + r_{j,k,h}^{(m,t)} \right)^2.
	\end{align}

	\item Step 3: Update the auxiliary variables $\{z_{j,k,h}^{(m,t+1)}\}$ by solving the following optimization problem:	
	\begin{equation}
	\label{eq-ADMM-zjk}
	\mathrm{argmin}_{\bm{z}_h} \quad
	\omega_h \frac{v}{2} \sum_{j=1}^p \sum_{k=1}^p s_{j,k} \left| z_{j,k,h} - z_{k,j,h} \right|
	+ \frac{\rho}{2} \sum_{j=1}^p \sum_{k=1}^p \left(z_{j,k,h} - \beta_{j,h}^{(m,t+1)} + r_{j,k,h}^{(m,t)} \right)^2.
	\end{equation}
	
	\item Step 4: Update the scaled dual variables $\{r_{j,k,h}^{(m,t+1)}\}$, which is equivalent to the residual between the $\beta_{j,h}^{(m,t+1)}$ and the auxiliary variable $z_{j,k,h}^{(m,t+1)}$. The estimates can be easily calculated by:
	\begin{equation}
	\label{eq-ADMM-rjk}
	r_{j,k,h}^{(m,t+1)} = r_{j,k,h}^{(m,t)} + \left(z_{j,k,h}^{(m,t+1)} - \beta_{j,h}^{(m,t+1)} \right).
	\end{equation}
	
	\item Step 5: If the termination conditions of the scaled ADMM algorithm are satisfied, the inner iteration will be stopped in advance,
	and they are taken as the estimates of the $(m+1)$th iteration of the EM update. Otherwise, the scaled ADMM algorithm will go back to Step 2 and continue until reaching the maximum number of steps $T$.
\end{itemize}
\vspace{-0.3em}

\emph{Remark} 1.
We use the standard stopping criterion for the scaled ADMM algorithm based on primal and dual residuals following Section 3.3 of \cite{boyd2011distributed}.
Specifically, we define the primal and dual residuals $\delta^{(m, t+1)}_{\text{pri},h}= ||\bm{r}_h^{(m,t+1)}-\bm{r}_h^{(m,t)}||_2, \ \delta^{(m, t+1)}_{\text{dual},h} = \rho  ||\bm{z}_h^{(m,t+1)}-\bm{z}_h^{(m,t)}||_2$ for the updates in each component, as well as the criterion $\delta^{(m, t+1)}_{\text{pri},h} \leq \varepsilon^{\text{pri}}$, $\delta^{(m, t+1)}_{\text{dual},h} \leq \varepsilon^{\text{dual}}$, where $\varepsilon^{\text{pri}}$ and $ \varepsilon^{\text{dual}}$ are specified as suggested by \cite{boyd2011distributed}.

\vspace{0.6em}
(3) Repeat E-step and M-step until the norm difference between the predicted regression coefficients of the two adjacent EM algorithms $\lVert \bm{\hat{\mathbf{B}}}^{(m+1)} - \bm{\hat{\mathbf{B}}}^{(m)} \lVert_{F}$ is less than the given threshold.

The EM-AMDD algorithm is implemented in R programming, as shown in Algorithm \ref{alg:EM-AMDD}.

\subsection{Computational details}\label{section-algorithm-details}
\begin{algorithm}
	\caption{EM-ADMM algorithm for FMR with covariate clustering.}
	\label{alg:EM-AMDD}
	\begin{algorithmic}[1] 
		\Statex
		\Require the observed data points $(y_1, \bm{x}_1), (y_2, \bm{x}_2), \dots, (y_n, \bm{x}_n)$, the initial estimation of the parameters $\bm{\Psi}^{(0)} = (\omega_1^{(0)},\dots,\omega_H^{(0)}, \beta_{0,1}^{(0)},\dots,\beta_{0,H}^{(0)}, \bm{\beta}_1^{(0)},\dots, \bm{\beta}_H^{(0)}, \phi_1^{(0)},\dots,\phi_H^{(0)})$, the penalty parameters $\gamma, v, \rho$, the maximum steps $M, T$, the threshold values $\varepsilon^{\text{pri}}, \varepsilon^{\text{dual}}, \varepsilon^{\text{EM}}$, and the similarity matrix $\mathbf{S}$;
		\Ensure the estimated $\bm{\Psi}^{*}$;
		
		\For{$m = 0,\dots,(M-1)$}
		\For{$h = 1,\dots,H$}
		\State calculate $\pi_{i,h}^{(m+1)}$ by \eqref{eq-EM-pi}-\eqref{eq-EM-pi-bx};
		\State update $\hat{\omega}_{h}^{(m+1)}$ by \eqref{eq-EM-weight};
		\State initialize $\bm{z}_h^{(m,0)} = \{z_{j,k,h}^{(m,0)} \}$ and $\bm{r}_h^{(m,0)} = \{r_{j,k,h}^{(m,0)} \}$ in the ADMM algorithm;
		\For{$t = 0,\dots,(T-1)$}
		\State update $\beta_{0,h}^{(m,t+1)}, \bm{\beta}_h^{(m,t+1)}, \phi_h^{(m,t+1)}$ by the BFGS algorithm and \eqref{eq-ADMM-beta0-deriv}-\eqref{eq-ADMM-betaj-deriv};
		\State update $\bm{z}_h^{(m,t+1)}$ by \eqref{eq-ADMM-zjk-update}-\eqref{eq-ADMM-zjk-theta};
		\State update $\bm{r}_h^{(m,t+1)}$ by \eqref{eq-ADMM-rjk};
		
		\If{($\delta^{(m, t+1)}_{\text{pri}} \leq \varepsilon^{\text{pri}} \ \And \ \delta^{(m, t+1)}_{\text{dual}} \leq \varepsilon^{\text{dual}}) \ \text{or} \ t=(T-1)$}
		\State $\hat{\beta}_{0,h}^{(m+1)} \leftarrow \beta_{0,h}^{(m,t+1)}$, $\hat{\bm{\beta}}_h^{(m+1)} \leftarrow \bm{\beta}_h^{(m,t+1)}$, $\hat{\phi}_h^{(m+1)} \leftarrow \phi_h^{(m,t+1)}$;
		
		\State stop;
		\EndIf
		\EndFor{\ ADMM iterations}
		\EndFor{\ all components}
		\\
		\If {$\lVert \bm{\hat{\mathbf{B}}}^{(m+1)} - \bm{\hat{\mathbf{B}}}^{(m)} \lVert_{F} < \varepsilon ^{\text{EM}} \ \text{or} \ m=(M-1)$}
		\State $\bm{\Psi}^{*} \leftarrow \bm{\hat{\Psi}}^{(m+1)}$;
		\State stop;
		\EndIf
		\EndFor{\ EM iterations}
		\State return $\bm{\Psi}^{*}$.
	\end{algorithmic}
\end{algorithm}

(1) Initialization of $\bm{\Psi}^{(0)}$.
Although numerical methods like EM and ADMM have good convergence properties for convex optimization problems, the choice of initial values remains important, particularly for problems with a large number of parameters to estimate. Using an uninformative prior can lead to suboptimal results, affecting the algorithm's convergence speed and stability.
Here, we propose a method that combines sample clustering with regularized GLM regression to address this issue.
The identification of subgroups of the outcome corresponds to the initial component weights of FMR models.
Specifically, we perform K-means clustering on observations $y_i$ to derive the prior weights of the components.
This approach is based on the idea that mixing is equivalent to classifying the latent risk level of each individual.
Using the clustering result, we partition the sample into $H$ subgroups and train a Ridge GLM model on each subgroup to obtain initial estimates of $\beta_{0,h}$, $\bm{\beta}_h$, and $\phi_h$, respectively.
This can provide a starting point for the subsequent EM or ADMM optimization.

(2) Selection of the tuning parameters.
The three tuning parameters in our model, $\gamma$, $v$, and $\rho$, each have a distinct role. Specifically, $\gamma$ controls the extent of shrinkage in the regression coefficients, $v$ regulates the penalty on the distance between similar covariates, and $\rho$ affects the convergence efficiency of the ADMM algorithm. For our study, we set $\rho=1$ and determined the optimal values of $\gamma$ and $v$ through cross-validation on the training set.
As we are primarily interested in the impact of the $\ell_2$ norm penalty and covariate clustering penalty, we evaluated the tuning parameter selection criteria using log-likelihood value or mean square error (MSE) criterion, given that our response variable of interest is continuous healthcare costs.
To provide insight into the computational requirements of our proposed EM-ADMM algorithm, we implemented the simulation setup outlined in Section \ref{section-simulation}, where $n=2000$ and $p=10$. On a computer equipped with a 3.61 GHz Intel Core-I7 CPU and 32 GB RAM, each EM-ADMM iteration took an average of 1.3 seconds to run. Matrix operations can be utilized to expedite the algorithmic speed.



(3) Identifying the covariates' clusters.
Although the objective function \eqref{eq-FMR-final-objective} theoretically guarantees that the difference between the coefficients of similar covariates will converge to zero, this result is difficult to achieve in practice due to limitations of numerical algorithms. Therefore, we follow the strategy proposed in \cite{chi2015splitting}, which places an edge between covariates $j$ and $k$ if $z_{j,k,h}$ is equal to $z_{k,j,h}$. Then, two covariates sharing the same edge can be classified into the same cluster in the $h$th component. This condition is established when $\theta = 0.5$ in \eqref{eq-ADMM-zjk-update}.

(4) Number of iterations and stopping criterion.
We set $M = 10$ and $T = 100$, which means that the EM algorithm is executed at most 10 times and each M-step involves a maximum of 100 ADMM iterations.
When determining the early termination conditions of the ADMM and EM algorithms, it is essential to consider the magnitude and dimension of the regression coefficients. In cases where the dimensionality of the covariates is high, we recommend setting the ADMM thresholds to $\varepsilon^{\text{pri}} = 0.05$ and $\varepsilon^{\text{dual}} = 0.05$, and the EM threshold to $\varepsilon^{\text{EM}} =0.01$.
We have found that these threshold values strike a balance between computational speed and convergence. This balance is achieved by monitoring the value of the log-likelihood function in subsequent numerical simulations and empirical studies.

\section{Simulation}\label{section-simulation}
A simulation study will be conducted to verify the effect of the proposed model on clustering and coefficient estimation for multiple groups of covariates.
The observations are assumed to be generated by a mixture distribution consisting of two components, where each component is a Gamma distribution with a mean parameter that depends on the covariates:
\[
Y_i \sim	\omega_1 \text{Gamma}(y_i \left| \mu_{1}(\bm{x}_i), \phi_1 \right. ) + \omega_2 \text{Gamma}(y_i \left| \mu_{2}(\bm{x}_i), \phi_2 \right. ),\quad i = 1,\dots, n,
\]
where $\mu_{h}(\bm{x}_i)$ and $\phi_h (h=1,2)$ respectively denote the mean and dispersion parameters, and $\omega_1 + \omega_2 = 1$.
The mean parameter $\mu_{h}(\bm{x}_i)$ can be expressed as a function related to the covariates $\bm{x}_i$ through a log link function,
specifically, $\log \mu_{h}(\bm{x}_i) = \beta_{0,h} + \bm{x}_i^T \bm{\beta}_h$.

The true values of the weights, regression coefficients and dispersion of the mixture Gamma model are set to be as follows:
\begin{align*}
	\text{Component} \ 1&: \omega_1=0.7, \
	\beta_{0,1}=1,\
	\bm{\beta}_1 = (\overbrace{\beta_{1,1},...,\beta_{5,1}}^{-0.1}, \overbrace{\beta_{6,1},...,\beta_{10,1}}^{-0.2}), \
	\phi_1 = 0.2;\\
\text{Component} \ 2&: \omega_2=0.3, \
\beta_{0,2}=2,\
\bm{\beta}_2 = (\overbrace{\beta_{1,2},...,\beta_{5,2}}^{0.1}, \overbrace{\beta_{6,2},...,\beta_{10,2}}^{0.2}), \
\phi_2 = 0.1;
\end{align*}
which is equivalent to dividing the 10 covariates of the $i$th observation $\bm{x}_i=(1, \overbrace{x_{i1},...,x_{i5}}^{\text{cluster 1}}, \overbrace{x_{i6},...,x_{i10}}^{\text{cluster 2}})$ into 2 clusters.
Each covariate in the same cluster shares a common effect on the outcome data.
For example, in component 1, the coefficient of covariate $k$ is 0.1 for $k=1,\dots,5$, and 0.2 for $k=6,\dots,10$. The covariates have completely opposite regression coefficients in the two components, reflecting the challenges caused by unobserved heterogeneous risk factors in healthcare cost modeling.

We generate the dataset with a sample size $n=1000$ during each run of the simulation.
The covariates $\bm{x}_i$ are assumed to be generated by a multivariate normal distribution $\text{MN}(\mathbf{0}, \mathbf{\Sigma})$ with the mean of $\mathbf{0}$ and the covariance matrix of $\mathbf{\Sigma}$.
$\mathbf{\Sigma}$ is a $p \times p$ block diagonal matrix, with the elements as follows:
\[
\Sigma_{kk'}=\begin{cases}
0.04 & \text{if}\quad k=k',\\
0.04\varrho & \text{if}\quad 1\le k\le 5, 1\le  k'\le 5, k\neq k',\\
0.04\varrho & \text{if}\quad 6\le k\le 10, 6 \le k'\le 10, k\neq k',\\
0& \text{otherwise},
\end{cases}
\]
where $\varrho$ denotes the correlation coefficient.
We can vary the value of $\varrho$ from the grid $\{0.1,0.5,0.9\}$ to explore the influence of covariate similarity on our model.
Thus, the observations $Y_i$ for $i=1,\dots, n$ are generated according to the following scenario: first, a random number $u_i$ is drawn from a uniform distribution over the interval $(0,1)$. If $u_i$ is less than $\omega_1$, $Y_i$ is drawn from the first component, specifically $\text{Gamma}(y_i \left| \mu_1(\bm{x}_i),\phi_1 \right.)$; otherwise, it is drawn from the second component, specifically $\text{Gamma}(y_i \left| \mu_2(\bm{x}_i),\phi_2 \right.)$.

\begin{figure}[htb]
	\centering
	\includegraphics[scale = 0.6]{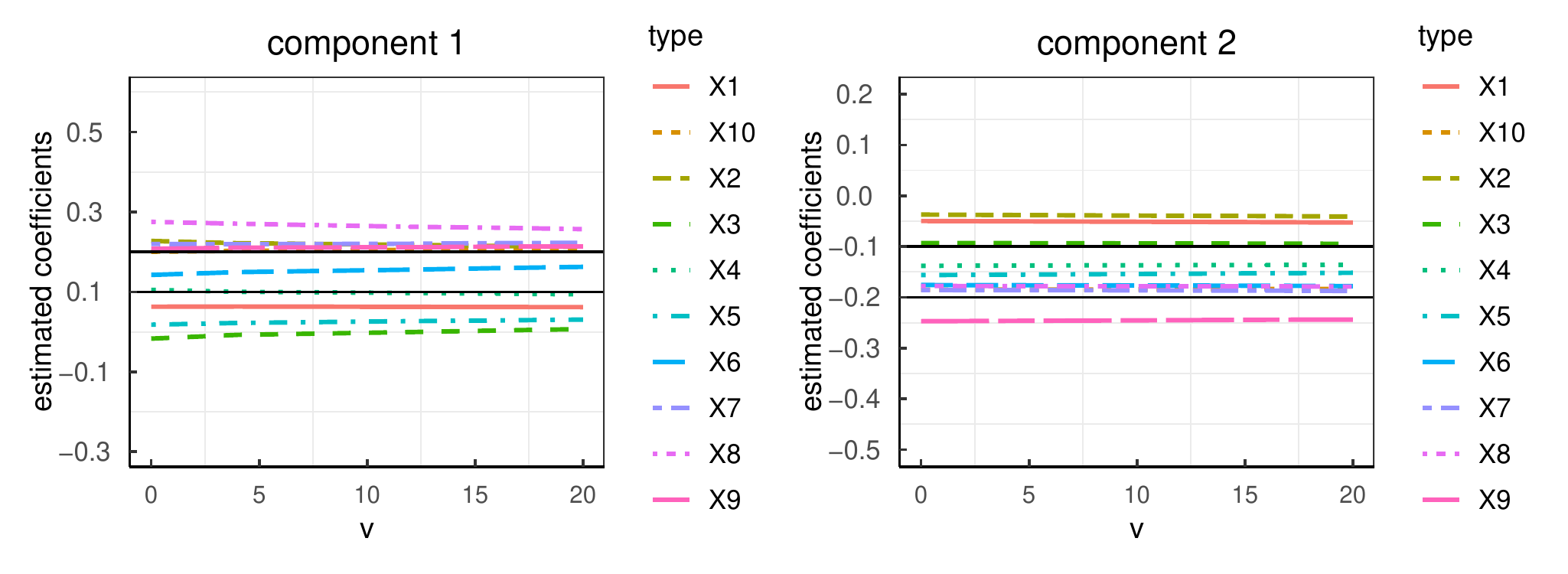}\\
	\includegraphics[scale = 0.6]{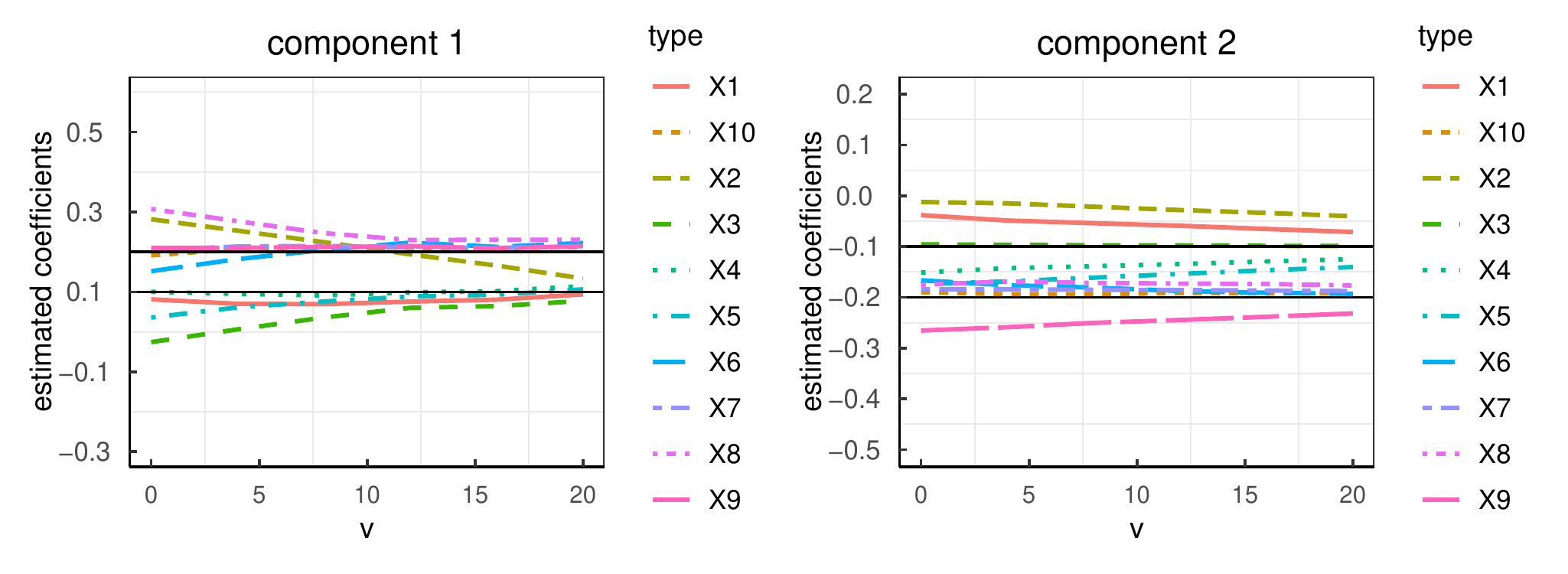}\\
	\includegraphics[scale = 0.6]{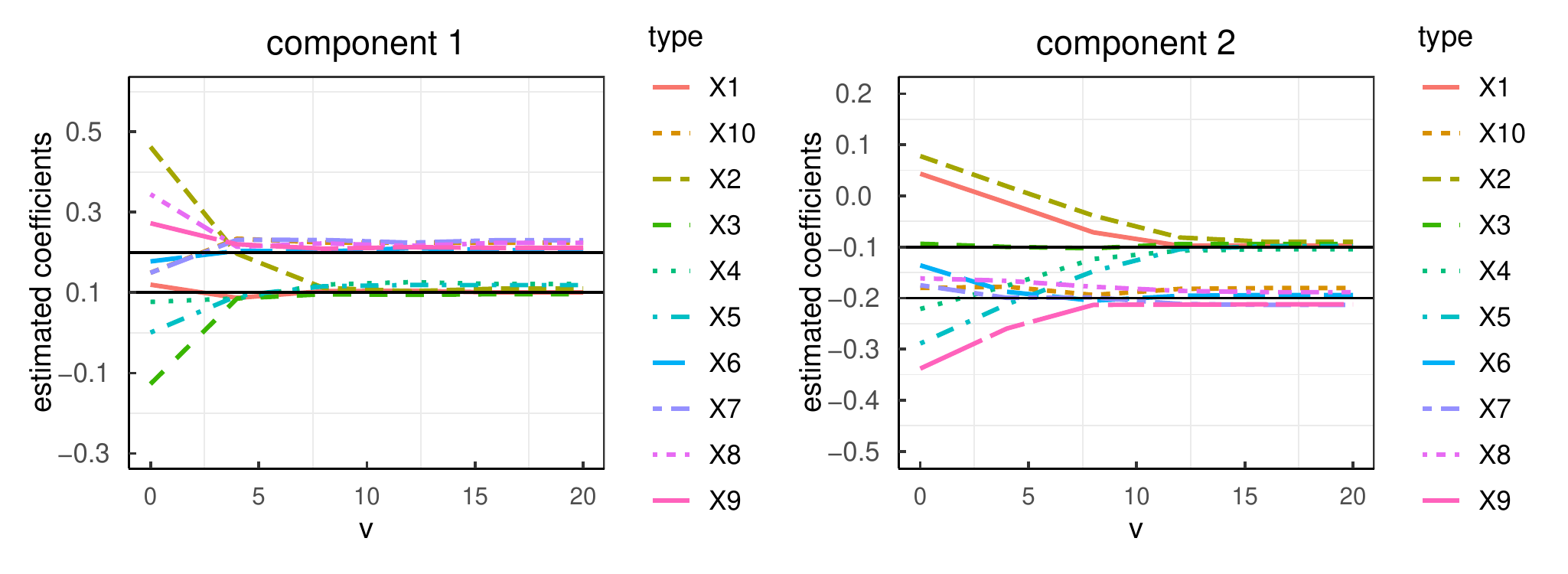}
	\caption{The solution paths for the covariates clustering against $v$ based on random samples with various values of $\varrho$. \textit{Left panel} and \textit{Right panel} display the estimated coefficients in component 1 and 2 respectively.
		The values of $\varrho=0.1,0.5,0.9$ are presented in the \textit{Top}, \textit{Middle}, and \textit{Bottom} panels, respectively.
		The true valuesof regression coefficients are marked as the black solid lines.
	}
	\label{fig-simulation-path}
\end{figure}

For each simulated dataset, we fit the proposed model with a wide range of tuning parameter values $v$\footnote{The other two tuning parameters are fixed at $\gamma=0.001$ and $\rho=1$.}. Figure \ref{fig-simulation-path} displays the solution paths against $v$ for clustering covariates under the three simulated samples of $\varrho = 0.1, 0.5, 0.9$. It is evident that if the effects of the covariates on the response variable exhibit different clustering effects in different subsamples, the proposed model will correctly cluster the covariates by identifying the same corresponding regression coefficients in each component. Our model also tends to cluster covariates when they have a certain correlation (e.g., $\varrho=0.5$), and it shows better identification of the main cluster with greater component weight in FMR. In Table \ref{tab: simulation-varrho}, we report the correctly clustered proportion (CCP) of the covariates among all observations of each latent component, which measures the agreement between the true and estimated clusters, and a value close to 1 is preferred. As expected, CCP-1 and CCP-2 both increase as the tuning parameter $v$ increases, especially when the true correlation of the covariates is high. The impact of covariate clustering is not apparent when the correlation is weak.

\begin{table}[htbp]
	\small
\centering
\caption{The correct clustering proportion (CCP) in each component for Gamma mixture regressions in the case that the covariates are assumed to be grouped into two clusters.}
\renewcommand\arraystretch{1.5}
\begin{tabular*}{\hsize}{@{}@{\extracolsep{\fill}}ccc|cc|cc@{}}
	\toprule
	\multirow{2}[0]{*}{$v$}
	& \multicolumn{ 2}{c|}{$\varrho=0.1$} & \multicolumn{ 2}{c|}{$\varrho=0.5$} & \multicolumn{ 2}{c}{$\varrho=0.9$} \\
\cline{2-7}
	\multicolumn{ 1}{c}{} &       CCP-1 &       CCP-2 &       CCP-1 &       CCP-2 &       CCP-1 &       CCP-2 \\
		\hline
	0 &    20.00\% &    20.00\% &    20.00\% &    20.00\% &    20.00\% &    20.00\% \\	
	4 &    40.00\% &    36.00\% &    64.00\% &    36.00\% &    52.00\% &    44.00\% \\
	8 &    44.00\% &    36.00\% &    88.00\% &    56.00\% &    88.00\% &    92.00\% \\
	12 &    44.00\% &    36.00\% &    92.00\% &    76.00\% &   100.00\% &   100.00\% \\
	16 &    44.00\% &    36.00\% &    96.00\% &    92.00\% &   100.00\% &   100.00\% \\
	20 &    44.00\% &    40.00\% &   100.00\% &   100.00\% &   100.00\% &   100.00\% \\
	\bottomrule
\end{tabular*}
	\label{tab: simulation-varrho}%
		\begin{tablenotes}
		\item * CCP-1 and CCP-2 denote the correct clustering proportion in component 1 and 2 respectively.
	\end{tablenotes}
\end{table}

We further conduct an investigation into the performance of the EM-ADMM algorithm in estimating parameters with respect to sample size $n$ and tuning parameter $v$, assuming high correlation of covariates in groups ($\varrho=0.9$).
The results of this evaluation are presented in Table \ref{tab: simulation-sample}, where bias and MSE were computed based on 100 repetitions for each of the sample sizes ($n=100,500,1000$).
Bias is defined as the difference between the average estimate and the true value of the regression coefficient, with a value close to zero indicating good performance.
MSE measures the average squared distance from the estimate to the true value, which is a commonly used metric for assessing prediction performance.
Table \ref{tab: simulation-sample} indicates that the model provides reasonable estimates for each coefficient, as long as a sufficient penalty parameter $v$ is added, regardless of the small or large sample size.
The accuracy of the parameter estimation results and the stability of the optimization algorithm are also reflected in the table. As the amount of training data increases, the estimation of the model becomes more accurate, and the magnitude of the bias and the MSE becomes almost negligible in comparison with the true values of the regression coefficients.

\begin{table}[htbp]
	\small
	\centering
	\caption{Bias and MSE for estimated regression coefficients for Gamma mixture models in the case of high correlation level of $\varrho = 0.9$.}
	\renewcommand\arraystretch{1.5}
	\begin{tabular*}{\hsize}{@{}@{\extracolsep{\fill}}lrrrrrrrrrrrrr@{}}
    	\toprule
&	& \multicolumn{4}{c}{$n=100$}     & \multicolumn{4}{c}{$n=500$}     & \multicolumn{4}{c}{$n=1000$} \\
	\cline{3-14}
&	     & \multicolumn{2}{c}{$v=0$} & \multicolumn{2}{c}{$v=20$} & \multicolumn{2}{c}{$v=0$} & \multicolumn{2}{c}{$v=20$} & \multicolumn{2}{c}{$v=0$} & \multicolumn{2}{c}{$v=20$} \\
&	& {Bias} & {MSE} & {Bias} & {MSE} & {Bias} & {MSE} & {Bias} & {MSE} & {Bias} & {MSE} & {Bias} & {MSE} \\
	\hline
    \multicolumn{1}{c}{\multirow{10}[0]{*}{Com1}} & $\beta_{1,1}$ & -0.031 & 0.045 & -0.003 & 0.000 & -0.005 & 0.005 & 0.001 & 0.000 & 0.001 & 0.004 & -0.001 & 0.000 \\
& $\beta_{2,1}$ & -0.021 & 0.047 & -0.002 & 0.000 & -0.006 & 0.006 & 0.000 & 0.000 & -0.003 & 0.003 & -0.001 & 0.000 \\
& $\beta_{3,1}$ & 0.023 & 0.055 & -0.001 & 0.000 & 0.010 & 0.007 & 0.000 & 0.000 & -0.004 & 0.003 & 0.000 & 0.000 \\
& $\beta_{4,1}$ & 0.013 & 0.051 & -0.002 & 0.000 & -0.004 & 0.007 & -0.001 & 0.000 & -0.010 & 0.003 & 0.000 & 0.000 \\
& $\beta_{5,1}$ & 0.006 & 0.033 & -0.003 & 0.000 & 0.003 & 0.005 & -0.001 & 0.000 & 0.013 & 0.003 & 0.000 & 0.000 \\
& $\beta_{6,1}$ & -0.001 & 0.046 & 0.000 & 0.000 & -0.003 & 0.008 & -0.002 & 0.000 & -0.007 & 0.003 & -0.001 & 0.000 \\
& $\beta_{7,1}$ & -0.021 & 0.041 & 0.000 & 0.000 & -0.016 & 0.005 & 0.000 & 0.000 & 0.009 & 0.003 & 0.002 & 0.000 \\
& $\beta_{8,1}$ & 0.001 & 0.038 & 0.000 & 0.000 & 0.007 & 0.005 & -0.002 & 0.000 & 0.002 & 0.003 & -0.001 & 0.000 \\
& $\beta_{9,1}$ & 0.012 & 0.037 & -0.001 & 0.000 & -0.001 & 0.005 & -0.001 & 0.000 & -0.009 & 0.003 & -0.002 & 0.000 \\
& $\beta_{10,1}$ & -0.001 & 0.036 & 0.002 & 0.000 & 0.008 & 0.005 & 0.000 & 0.000 & 0.001 & 0.003 & -0.001 & 0.000 \\
\hline
\multirow{10}[0]{*}{Com2} & $\beta_{1,2}$ & -0.082 & 0.535 & -0.003 & 0.002 & 0.022 & 0.078 & -0.006 & 0.000 & 0.012 & 0.028 & 0.000 & 0.000 \\
& $\beta_{2,2}$ & -0.044 & 0.344 & -0.002 & 0.002 & -0.034 & 0.053 & -0.005 & 0.000 & 0.028 & 0.030 & 0.001 & 0.000 \\
& $\beta_{3,2}$ & 0.060 & 0.475 & -0.003 & 0.002 & 0.019 & 0.044 & -0.005 & 0.000 & -0.025 & 0.030 & 0.000 & 0.000 \\
& $\beta_{4,2}$ & 0.053 & 0.427 & -0.003 & 0.002 & -0.028 & 0.051 & -0.006 & 0.000 & -0.012 & 0.029 & -0.001 & 0.000 \\
& $\beta_{5,2}$ & -0.031 & 0.375 & 0.000 & 0.002 & -0.012 & 0.064 & -0.005 & 0.000 & -0.011 & 0.027 & -0.002 & 0.000 \\
& $\beta_{6,2}$ & 0.024 & 0.396 & -0.011 & 0.003 & -0.006 & 0.052 & -0.004 & 0.000 & 0.007 & 0.023 & 0.001 & 0.000 \\
& $\beta_{7,2}$ & -0.007 & 0.445 & -0.014 & 0.003 & -0.007 & 0.048 & -0.004 & 0.000 & 0.004 & 0.026 & 0.001 & 0.000 \\
& $\beta_{8,2}$ & -0.048 & 0.415 & -0.012 & 0.003 & -0.007 & 0.059 & -0.003 & 0.000 & -0.004 & 0.026 & -0.001 & 0.000 \\
& $\beta_{9,2}$ & -0.099 & 0.505 & -0.012 & 0.002 & -0.003 & 0.052 & -0.003 & 0.000 & -0.025 & 0.027 & 0.000 & 0.000 \\
& $\beta_{10,2}$ & 0.053 & 0.375 & -0.012 & 0.003 & 0.002 & 0.053 & -0.003 & 0.000 & 0.014 & 0.027 & 0.001 & 0.000 \\
\bottomrule
\end{tabular*}%
	\label{tab: simulation-sample}%
\end{table}

\clearpage
\section{Empirical analysis}\label{section-empirical}
\subsection{Experimental setup}
In this section, we present two practical examples to demonstrate the proposed covariate clustering regression model. The first example uses a dataset on diabetes expenses from several hospitals in China, which includes patient characteristics and questionnaire data. The aim is to investigate the relationship between annual expenses and individuals' knowledge of diabetes. The second example uses a short-term medical claim dataset collected from a Chinese insurance company. The objective is to explore the clustered effect of the ICD (International Classification of Diseases) categories in the subgroups of the medical insurance claim data. To prevent model overfitting, we divide the dataset $\mathcal{L}(y_{i},\bm{x}_i){0\le i\le n}$ into a training set, which is used for model fitting and hyperparameter tuning, and a testing set, which is used for out-of-sample analysis only.

A comparative study is conducted for model comparison in terms of goodness of fit, predictive accuracy, and discriminative ability of samples of different risk in the testing dataset. Two commonly used metrics to evaluate the goodness of fit of a model are the negative log-likelihood (NLL) and the pseudo R$^2$.
The Gamma distribution is chosen as the underlying distribution assumption for all models, so the density function $f(y)$ in \eqref{eq-FMR-loglike} has the form
\begin{equation}
f\left(y \left| \mu, \phi \right. \right) =
\frac{y^{\frac{1}{\phi^2}-1} \exp\left(-\frac{y}{\mu\phi^2}\right)} {(\mu\phi^2)^{\frac{1}{\phi^2}} \ \Gamma({1}/{\phi^2})},
\end{equation}
where $\Gamma(\alpha) = \int_{0}^{\infty} t^{\alpha-1}e^{-t}dt$ is the gamma function, and NLL can be defined as $-\ell_n(\bm{\Psi})$ with a set of estimated parameters.
The pseudo R$^2$ measures the gap of log-likelihood between the saturated and predictive model.
Specifically, given the sample observations $\bm{y} = ({y}_1, \dots, {y}_n)$ and the predictions $\hat{\bm{y}} = (\hat{y}_1, \dots, \hat{y}_n)$, where $\hat{y}_i = \sum_{h=1}^H \hat{\omega}_h \hat{\mu}_h(\bm{x}_i)$,
the pseudo R$^2$ under the Gamma distribution assumption can be expressed as:
\begin{equation}
\text{R}^2 = 1 - \frac{\sum_{i=1}^n \left[\log(y_i / \hat{y}_i) - (y_i-\hat{y}_i)/\hat{y}_i \right]} {\sum_{i=1}^n \log(y_i / \bar{y})},
\end{equation}
where $\bar{y}$ is the sample mean of $\bm{y}$. A low value of NLL and a high value of R$^2$ are preferred for the model.
Regarding predictive accuracy, we employ two metrics: mean squared error (MSE) to evaluate the accuracy of point estimation, and mean continuous ranked probability score (MCRPS) to evaluate the accuracy of distribution estimation.
While MSE is easy to calculate, CRPS measures how well the overall predicted distribution takes values around the true observation. Its analytical form is given by \cite{gneiting2007strictly} as:
\begin{equation}
\text{CRPS}(y_i, \hat{F}) = \int_{\mathbb{R}} \left[\hat{F}(y_i|\bm{x}_i) - \text{I}(y_i \leq z) \right]^2 dz,
\end{equation}
{where $\hat{F}(\cdot)$ is the estimated cumulative distribution function corresponding to \eqref{eq-FMR-density}, and $\text{I}(\cdot)$ is an indicative function.}
A predicted distribution with a smaller CRPS would logically be more concentrated around the true observation. We then average the CRPS values of all samples to obtain the model's performance on the entire dataset. Finally, we use the two-way lift metric to assess the model's ability to discriminate risk across samples. To calculate the lift, we bin the samples into ten equal groups according to their predicted losses, from low to high. We then divide the average response of the top-decile group by that of the bottom-decile group \citep{lee2021addressing}. A higher lift means that the model can better distinguish between high-risk and low-risk samples.

\subsection{Diabetes expense dataset}
The first dataset comprises healthcare costs data for 5,961 diabetes patients collected from several large hospitals in China during the year 2018.
Each observation records the positive annual expense of a patient that covers treatment and hospitalization costs incurred due to diabetes.
The dataset contains patient-specific covariates related to the healthcare costs, primarily demographic characteristics such as gender, age, monthly income, payment method,
and basic health status such as smoking, drinking, duration of diabetes, body mass index, checkup values of blood pressure and blood glucose.
The dataset also includes information on complications of diabetes, such as chronic complications (including hypertension and vessel disease) and acute complications, as well as medication use, including whether oral hypoglycemic drugs or insulin are used.
These variables are aggregated to form 28 traditional risk factors.
Additionally, a questionnaire survey is conducted for each patient to evaluate their knowledge of diabetes, which consists of 26 main multiple-choice questions, with a total of 111 options. The questionnaire survey records whether each option is answered correctly, denoted as binary variables of 0 and 1.


We aim to investigate the association between the individuals' knowledge of diabetes and diabetes healthcare cost. 
However, directly modeling the original 111 covariates is not advisable, as the information reflected by different options of each question, or even different questions, may overlap to some extent. Consequently, we will only cluster the 111 covariates of the questionnaire in the follow-up experiments and retain the traditional covariates without further processing. Synchronized covariate clustering in regression modeling is a viable technique for reducing dimensionality and mining the textual information of the questionnaire. We adopt a data-driven approach to construct the similarity matrix $\mathbf{S}={s_{j,k}}_{p \times p}$ by computing the cosine similarity of covariate $j$ and covariate $k$ in a training set, while more sophisticated methods of defining covariate similarity are worth exploring in future research.

\begin{figure}[htb]
	\centering
	\includegraphics[scale = 0.7]{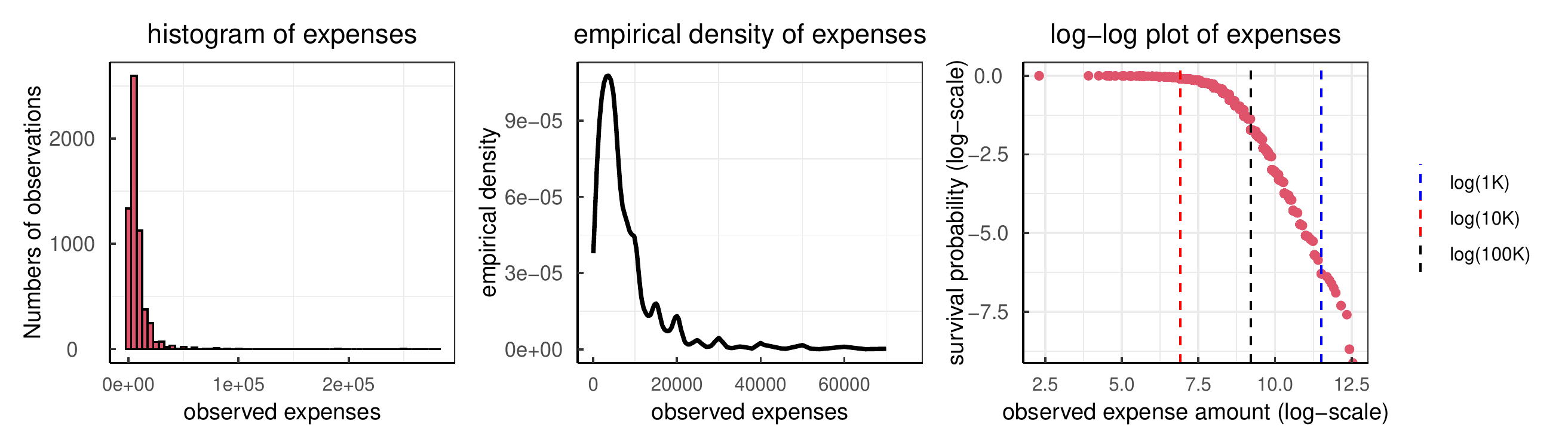}
	\caption{Diabetes expense dataset: \textit{Left panel}: histogram of the patients' healthcare costs;
		\textit{Middle panel}: empirical density (upper-truncated at 75,000);
		\textit{Right panel}: log-log plot of observed expenses.}
	\label{fig-diabetes-dist}
\end{figure}

Table \ref{tab-diabetes-summary} provides an overview of the 139 covariates detailing the risk factors.
Figure \ref{fig-diabetes-dist} illustrates the empirical distribution of diabetes expenses. The log-log plot indicates that the largest 10\% of the expense tail exhibits a linear trend, implying that the response variable is heavy-tailed. Additionally, the density function is highly skewed and, to some extent, multi-modal; however, the subsequent peaks are not significant.

\begin{table}[htbp]
	\small
	\centering
	\caption{Diabetes expense dataset: description of variables.}
	\renewcommand\arraystretch{1.5}
	\begin{tabular*}{\hsize}{@{}@{\extracolsep{\fill}}lll@{}}
		\toprule
		\textbf{Variable} & \textbf{Type} & \textbf{Description}\\
		\hline
		healthcare costs& Continuous& Annual healthcare costs of an inpatient for diabetes: 3-200,000 CNY.\\
		
        Gender& Categorical& Binary: male or female.\\
		Age& Continuous& Integer number: 16-94 years.\\
		Education level& Categorical& Six types: from below elementary school to undergraduate and above.\\
        Marital status& Categorical& Binary: married or not.\\
        Occupation& Categorical & Four types: on-the-job, retired, laid-off, other.\\
        Income& Continuous& Monthly income of the patient: 24-100,000 CNY.\\
		Pay way& Categorical& \tabincell{l}{Payment way of healthcare costs, four types: self-pay, public-pay,\\ medical insurance, other.}\\
		
        Smoking& Categorical& Binary: smoking or not.\\
		Drinking& Categorical& Binary: drinking or not.\\
		Duration of diabetes& Continuous& How long since diabetes was diagnosed: 1-84.7 years.\\
		Physical indicators& Continuous& \tabincell{l}{A total of 6 covariates, reflecting the basic health status of the inpatient.\\
            These covariates include: body mass index (12.80-60.44), systolic blood\\ pressure (60-240mmHg), diastolic blood pressure (30-180mmHg), fasting\\ plasma glucose (0.56-43.20mmol/L), 2-hour postprandial glucose (0.9-55\\mmol/L), and hemoglobin A1c (3-18mmol/L). All are continuous.}\\

        Chronic complications& Categorical& \tabincell{l}{A total of 7 covariates, recording whether the inpatient has chronic\\ complications of diabetes (i.e.,
            hypertension, vessel disease, diabetic\\ nephropathy, diabetic neuropathy, diabetic eye disease, diabetic foot\\ disease, and others). All are binary variables,
            with ``1'' being diseased.}\\
		Acute complications& Categorical& \tabincell{l}{A total of 3 covariates, recording whether the inpatient has acute comp-\\lications of diabetes (i.e.,
            diabetic ketoacidosis, hypoglycemia, and lactic\\ acidosis). All are binary variables, with ``1'' being diseased.}\\

        Oral medication& Categorical& Binary: whether oral hypoglycemic drugs are used or not.\\
        Insulin medication& Categorical& Binary: whether insulin is used or not.\\

        Questionnaire data& Categorical& \tabincell{l}{A total of 111 covariates, measuring the individual's understanding of\\ diabetes-related knowledge.
            All are binary variables: a certain covariate\\ is recored as ``1'' if the corresponding option is answered correctly.}\\
		\bottomrule
	\end{tabular*}
	\label{tab-diabetes-summary}
\end{table}

Table \ref{tab-diabetes-comparison} shows the out-of-sample performance of competing models, covering a comprehensive comparison among conventional GLMs, ridge GLMs and covariate clustering GLMs, as well as the corresponding mixture models.
Specifically,
M1 to M4 represent the conventional GLM (without $\ell_2$-penalty and similarity penalty), the ridge GLM (without similarity penalty), the mixture GLM (without $\ell_2$-penalty and similarity penalty), and the mixture ridge GLM (without similarity penalty), respectively.
M5-M8 are similar to M1-M4, but the last four models all incorporate our covariate clustering framework (i.e., with similarity penalty).
To speed up the convergence of the algorithm, the response variable is transformed into $y^*=y/10,000$, all continuous covariates are normalized, and the other (multiple) categorical covariates are treated with dummy variables.
The dataset is randomly split into the training and testing sets with a proportion of 80:20, leading to 4,768 observations for training and 1,193 observations for testing.
Regarding tuning parameters selection, when the $\ell_2$ norm regularization term is considered, $\gamma$ is tuned from the three values ${0.5, 1.0, 5.0 }$; otherwise, $\gamma = 0.01$ is fixed to ensure the strict convexity of the objective function.
When the regularization term for covariate clustering is added, $\rho = 1.0$ is fixed, and $v$ is tuned from the grid values $\{0.5, 1.0, \dots, 4.5, 5.0 \}$; otherwise, $\rho = 0.001$, and $v = 0$ are fixed. We select the tuning parameters that can achieve the minimum NLL on the validated set for each model through further subdividing a validated part from the training set.

\begin{table}[htbp]
	\centering
	\caption{Diabetes expense dataset: out-of-sample performance of the eight competing models.}
	\label{tab-diabetes-comparison}
	\begin{tabular}{cccccccccc}
		\toprule
		\multirow{2}[0]{*}{Models} & \multicolumn{4}{c}{Hyperparameters} & \multicolumn{5}{c}{Out-of-sample Performance} \\
		\cmidrule(lr){2-5} \cmidrule(lr){6-10} & H & $\gamma$ & $\rho$ & $v$ & NLL & Pseudo R$^2$ & MSE(10,000) & MCRPS & Lift \\
		\hline
		M1 & 1 & 0.01 & 0.001 & 0 & 733.718 & 15.83\% & 1.384 & 0.363 & 3.588 \\
		M2 & 1 & 0.5  & 0.001 & 0 & 733.743 & 15.83\% & 1.384 & 0.363 & 3.588 \\
		M3 & 2 & 0.01 & 0.001 & 0 & 801.783 & 11.97\% & 1.373 & 0.356 & 3.716 \\
        M4 & 2 & 5.0  & 0.001 & 0 & 812.429 & 11.08\% & 1.383 & 0.357 & 3.606 \\
        \hline
		M5 & 1 & 0.01 & 1.0 & 0.5 & \textbf{716.809} & \textbf{17.51\%} & 1.388 & 0.359 & \textbf{4.250} \\
		M6 & 1 & 1.0  & 1.0 & 0.5 & 717.408 & 17.44\% & 1.389 & 0.359 & 4.210 \\
		M7 & 2 & 0.01 & 1.0 & 3.5 & 780.983 & 14.04\% & \textbf{1.368} & \textbf{0.354} & 3.723 \\
        M8 & 2 & 1.0  & 1.0 & 2.5 & 793.535 & 12.29\% & 1.376 & 0.354 & 4.064 \\
		\bottomrule
	\end{tabular}
	\begin{tablenotes}
	\item *The optimal values of metrics are in bold.
\end{tablenotes}
\end{table}

Based on the results presented in Table \ref{tab-diabetes-comparison}, several main conclusions can be drawn. Firstly, the application of the covariate clustering method significantly improves the predictive performance of diabetes costs in this small dataset characterized by a large number of highly correlated risk factors. The comparison between models M5 to M8 and models M1 to M4 reveals that most of the metrics of the former set are superior to those of the latter, with the exception of the MSEs of M5 and M6, which are slightly worse than those of M1 and M2. Notably, the covariate clustering model not only accurately estimates the expected value (point estimates) but also effectively characterizes the overall distribution of the target outcome, as evidenced by the improvement in goodness of fit (i.e., NLL and pseudo R$^2$) and the ability to discriminate extreme observations (i.e., lift).
This phenomenon is consistent with our expectations that the covariate clustering method is equivalent to aggregating the effects of multiple related risk factors, which eliminates the random disturbance and mutual interference among the covariates, ultimately enhancing the data.

Secondly, once the covariates with similar or repeated effects have been clustered during the regression process, the original $\ell_2$ norm penalty for the compression and selection of covariates is no longer necessary. The results indicate that the performance of models {M2, M4, M6, M8} is not as optimal as that of models {M1, M3, M5, M7}. As previously mentioned, dimensionality reduction can also be achieved by clustering similar covariates, rendering the use of the $\ell_2$ norm penalty unnecessary in subsequent experiments.

Thirdly, when comparing mixture and non-mixture models, model M5 exhibits the best performance in terms of goodness of fit and risk discrimination ability, while M7 performs best in terms of prediction accuracy. Although non-mixture models can effectively capture the characteristics of diabetes expenses data where the heterogeneity is insignificant, mixture models can generally capture the majority values of the sample and have advantages in point estimation.

\begin{figure*}[!htb]
	\centering
	\subfloat[covariate clustering results in component 1 \label{fig-3a}] {\includegraphics[width=0.48\textwidth] {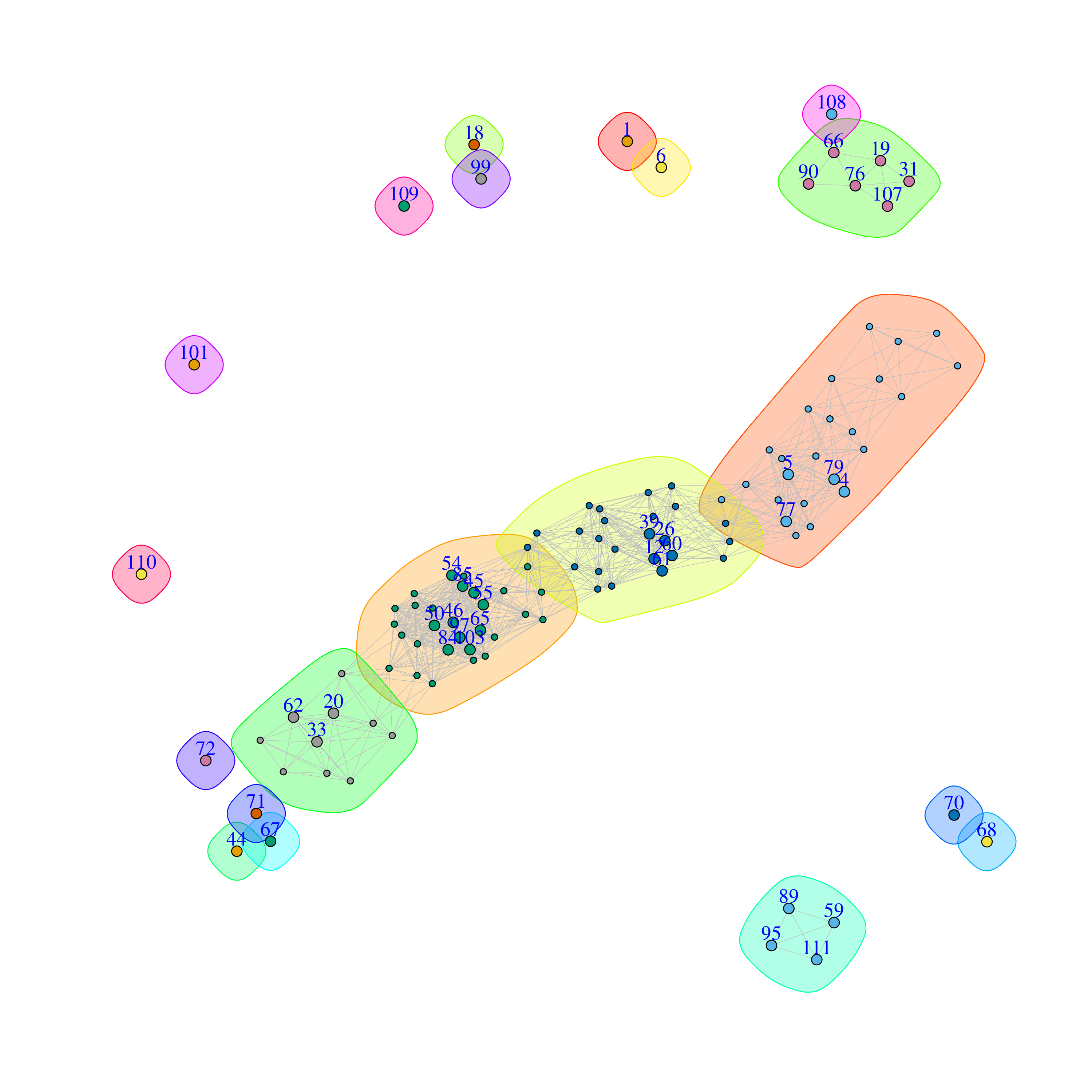}} \quad
	\subfloat[covariate clustering results in component 2 \label{fig-3b}] {\includegraphics[width=0.48\textwidth] {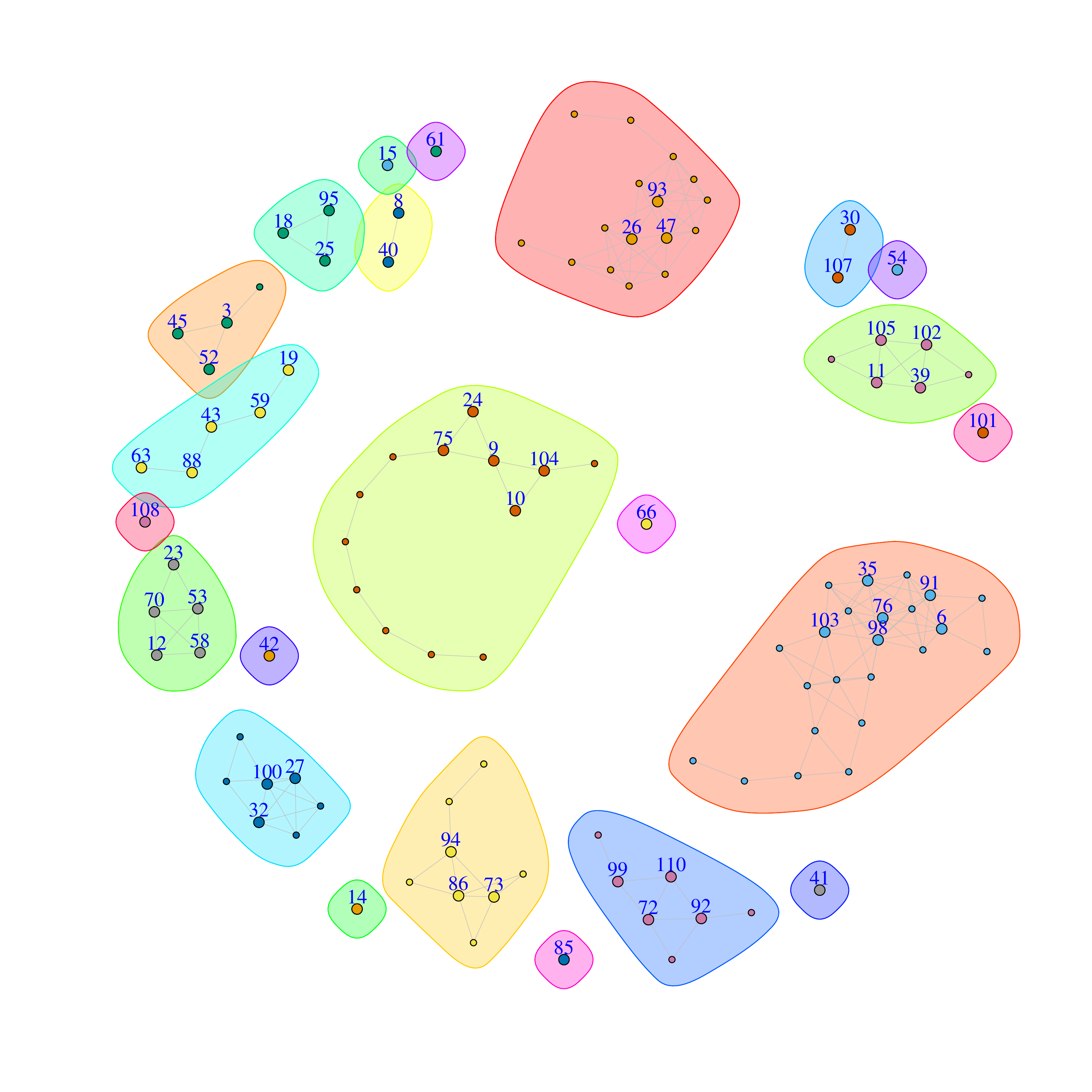}} \\
	\caption{Diabetes expense dataset: clustering results of the proposed model: the vertex is a covariate, the edge is an association between two covariates, and the covariates can be grouped into a single cluster if a connected path exists between them; the colored background depicts a covariate's ``community'' with finer granularity than the ``cluster''.}
	\label{fig-diabetes-cluster-results}
\end{figure*}

As demonstrated previously, our proposed framework effectively integrates prior similarity with model performance to generate meaningful posterior clusters and regression coefficients for the covariates. We employ network analysis to examine the relationships between the covariates in the resulting model.
Figure \ref{fig-diabetes-cluster-results} depicts the unweighted and undirected graphs. It is worth noting that the distance between non-connected clusters does not reflect their similarity. In component 1, the covariates are predominantly grouped into one main cluster (located in the middle) and two minor clusters (in the upper-right and lower-right regions), while the remaining covariates exhibit unique effects that cannot be aggregated. Component 2, however, yields more distinct clusters, allowing for better differentiation of the covariates' effects. This finding is in line with the prior similarity analysis.
In other words, the model clustering most covariates in subgroup 1 (as in component 1, which accounts for approximately $\hat{\omega}_1=0.96$ of the mixture model's weight),
while it fine-tunes the clustering effects on subgroup 2 (as in component 2, $\hat{\omega}_2=0.04$).

Furthermore, the utilization of colored backgrounds provides a more nuanced representation of clustering outcomes, where community formation is determined by the edge betweenness of each vertex \citep{girvan2002community}. Moreover, we have identified and compiled a list of significant covariates for large communities, enabling further exploration of their internal relationships. For instance, component 1's largest cluster can be subdivided into four communities, with one of them containing ``V4" ,``V5" ,``V77", and ``V79," which relate to fundamental knowledge and the detrimental effects of smoking on diabetes patients.
The exhaustive description of these variables is available in Appendices \ref{section-appC}.

The comprehensive findings underscore the limitations of implementing unsupervised clustering and the conventional regression methods in isolation. 
Our framework concurrently addresses both tasks to attain more precise results for both regression and covariate clustering. While we have solely conducted elementary analyses based on the covariates' network structure, adopting more semantic methods in the future can enhance the similarity matrix's definition and yield more comprehensive outcomes regarding patients' diabetes self-awareness. 

\subsection{Short-term medical claim dataset}
As a second example, we examine a medical insurance dataset provided by a major insurance company in China, which includes individual inpatient claim data for a short-term medical insurance plan covering the period from 2014 to 2016. The dataset comprises 19,110 policies, and each policy includes the positive inpatient claim amount as well as several covariates such as \textit{Gender}, \textit{Age}, \textit{SSCoverage}, \textit{HospitalDays}, \textit{ClaimType}, and \textit{International Classification of Diseases (ICD) code}. Table \ref{tab-medical-summary} displays the summary statistics of these variables.

\begin{figure}[htb]
	\centering
	\includegraphics[scale = 0.5]{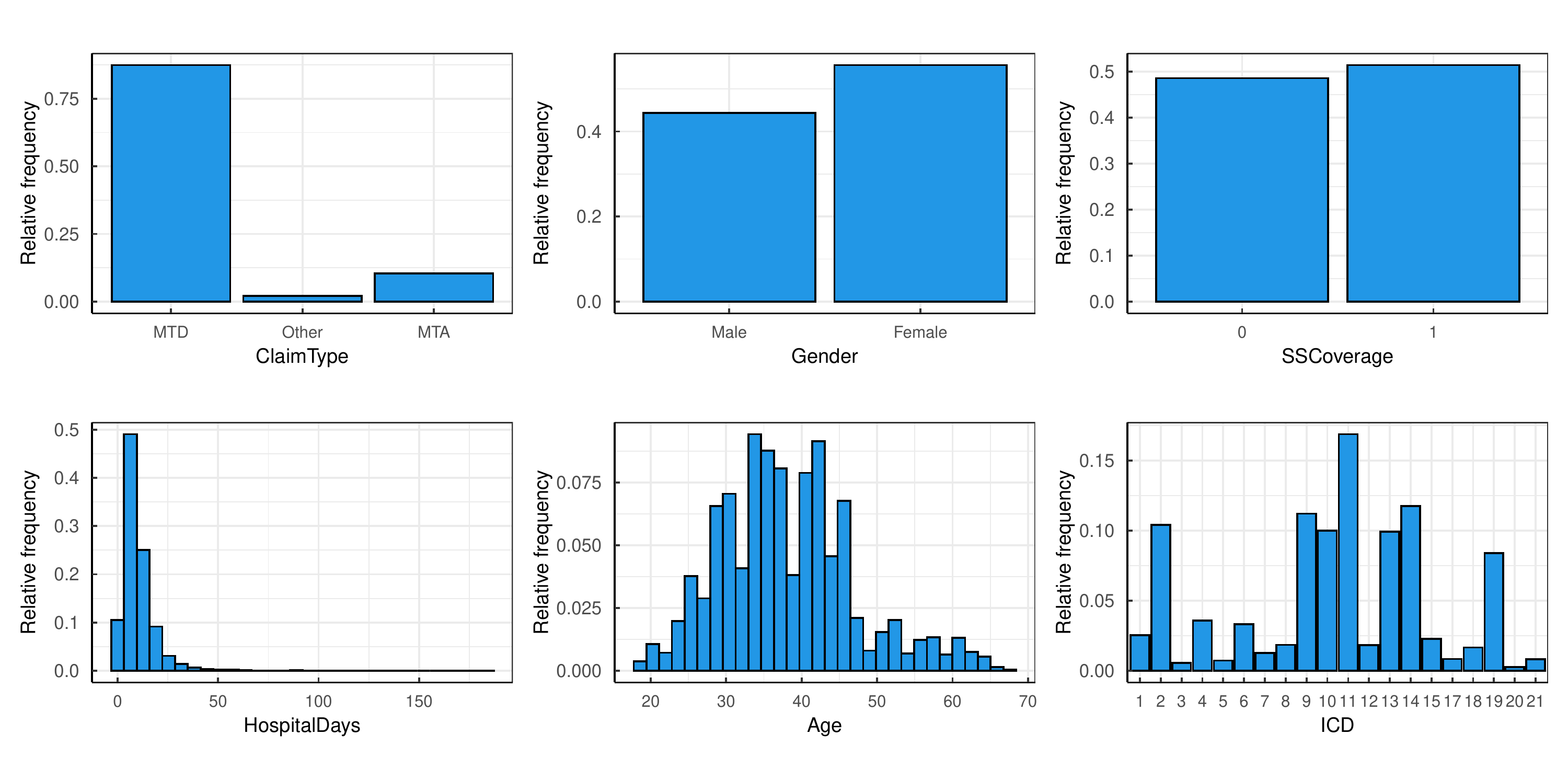}
	\caption{Medical claims case: relative frequency of the covariates \textit{Gender}, \textit{Age}, \textit{SSCoverage},
		\textit{ICD},
		 \textit{HospitalDays} and \textit{ClaimType}.}
	\label{fig-medical-frequency}
\end{figure}

Figure \ref{fig-medical-frequency} shows how the covariates are distributed in the medical insurance dataset.
We model the claim amounts by incorporating covariates through the two-component mixture Gamma regression models for covariates clustering,
illustrating how clustering the ICD dummy variables can improve the medical claims prediction.
For our analysis, we split the dataset into a training set and a testing set in proportion 70:30.
To increase the convergence of the algorithm,  two continuous covariates (\textit{Age} and \textit{HospitalDays}) are standardized.
The ICD variables are transformed into 20 dummy variables.
We consider the non-informative similar matrix for these ICD dummy variables by setting $S_{kk'} = 0.5$ for $k,k'\in\left\{1,\dots,20\right\}$.
For non-similarity covariate $i\ne k$ and $j \ne k'$, we set $S_{ij}=0$.
The regression coefficients are estimated by using proposed EM-ADMM with $\rho=1$ and $\gamma= 0.01$.
The optimal tuning parameter for covariate clustering is $v=4.5$, which is selected from $v \in \left\{0, 0.5, 1, \dots, 5\right\}$ based on the minimum NLL criteria on training data.

\begin{figure}[htb]
	\centering
	\includegraphics[scale = 0.6]{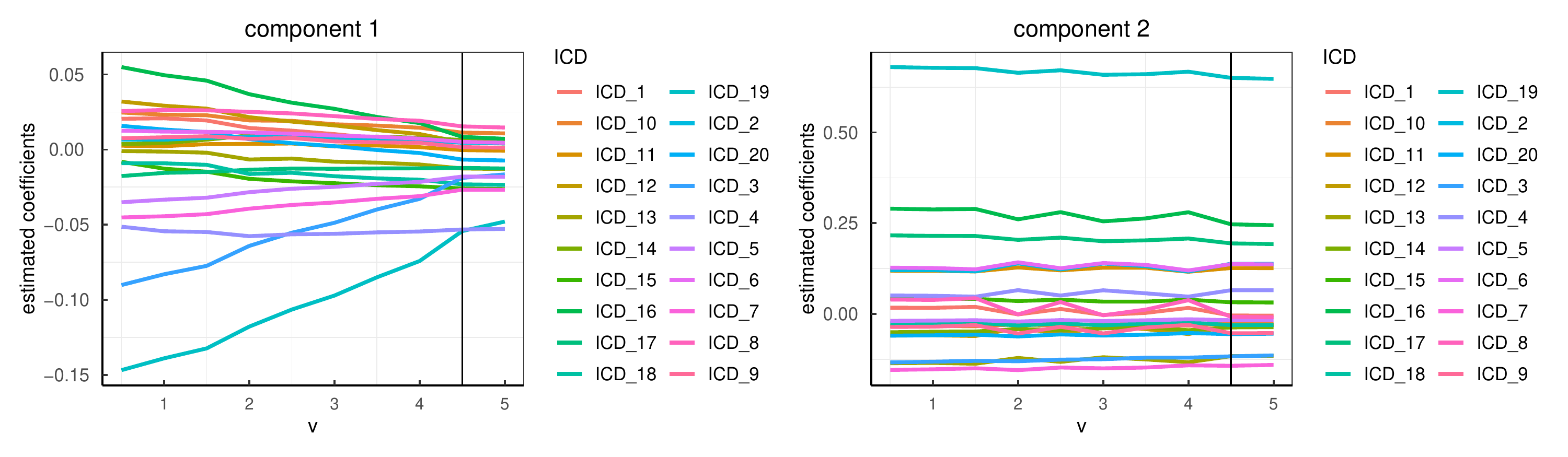}
	\caption{Medical claim dataset: the estimated regression coefficients for two components in mixture Gamma regression model when $v$ varies from 0 to 5 in the training dataset.
		The optimal $v=4.5$ is marked as the vertical line. The estimated weights for the two components are $\hat{w}_1=95.33\%$ and  $\hat{w}_2= 4.67\%$ respectively.
}
	\label{fig-medical-path}
\end{figure}

Figure \ref{fig-medical-path} shows the estimated coefficients for two components respectively in mixture regression model with different values of $v$.
An ideal covariate clustering method should achieve high prediction accuracy with few clusters, as fewer clusters lead to more compact models and potentially easier interpretability.
We find that the number of clustering groups for the ICD covariates in the first second component tends to decrease as the tuning parameter $v$ increases, indicating that the covariate clustering method for mixture regression leads to significantly more compact models than the mixture model without covariate clustering, especially in the first component.
Specifically, the proposed method makes it possible to classify ICD18 and ICD4 into one category by taking into account the outcome's information.
However, since the weight parameter in the second component is estimated for the range of $w_2\in \left[0, 0.05\right]$, which gives less weight to the penalty of the log-likelihood function in the second component, the proposed method has different performance in each subject subgroup.

\begin{table}[htbp]
	\small
	\centering
	\caption{Medical claim dataset: description of variables.}
	\renewcommand\arraystretch{1.5}
	\begin{tabular*}{\hsize}{@{}@{\extracolsep{\fill}}lll@{}}
		\toprule
		\textbf{Variables} & \textbf{Type} & \textbf{Description}\\
		\hline
		Claim amount&	Continuous	&inpatient (positive) claim amount of a patient \yen 3-200,000 \\
		Age&	Continuous&	inpatient's age: 18-67\\
		Gender &	Categorical & gender of a inpatient (2 levels): male, female\\
		SSCoverage &	Categorical & social medical insurance or public medical insurance coverage (2 levels): yes or no\\
		HospitalDays &	Continuous & the duration of hospitalization of a inpatient: 0-184 days\\
		ClaimType &	Categorical & \tabincell{l}{reasons for claiming the medical insurance claim (3 levels):\\ medical treatment from disease (MTD), \\ medical treatment from accident (MTA), other}\\
		ICD & Categorical & \tabincell{l}{ICD-10 (20 levels): \\
	1: A00-B99, Certain infectious and parasitic diseases\\
	2: C00-D49, Neoplasms\\
	3: D50-D89, Diseases of the blood and blood-forming organs and certain disorders \\
	\quad \quad \quad \quad \quad \quad involving the immune mechanism\\
	4: E00-E89, Endocrine, nutritional and metabolic diseases\\
	5: F01-F99, Mental, Behavioral and Neurodevelopmental disorders\\
	6: G00-G99, Diseases of the nervous system\\
	7: H00-H59, Diseases of the eye and adnexa\\
	8: H60-H95, Diseases of the ear and mastoid process\\
	9: I00-I99, Diseases of the circulatory system\\
	10: J00-J99, Diseases of the respiratory system\\
	11: K00-K95, Diseases of the digestive system\\
	12: L00-L99, Diseases of the skin and subcutaneous tissue\\
	13: M00-M99, Diseases of the musculoskeletal system and connective tissue\\
	14: N00-N99, Diseases of the genitourinary system\\
	15: O00-O9A, Pregnancy, childbirth and the puerperium\\
	17: Q00-Q99, Congenital malformations, deformations and chromosomal abnormalities\\
	18: R00-R99, Symptoms, signs and abnormal clinical and laboratory findings,\\
	 \quad \quad \quad \quad \quad \quad  not elsewhere classified\\
	19: S00-T88, Injury, poisoning and certain other consequences of external causes\\
	20: V00-Y99, External causes of morbidity\\
	21: Z00-Z99,  Factors influencing health status and contact with health services\\
}
		\\
		\bottomrule
	\end{tabular*}
	\label{tab-medical-summary}
\end{table}


\begin{figure}[htb]
	\centering
	\includegraphics[scale = 0.6]{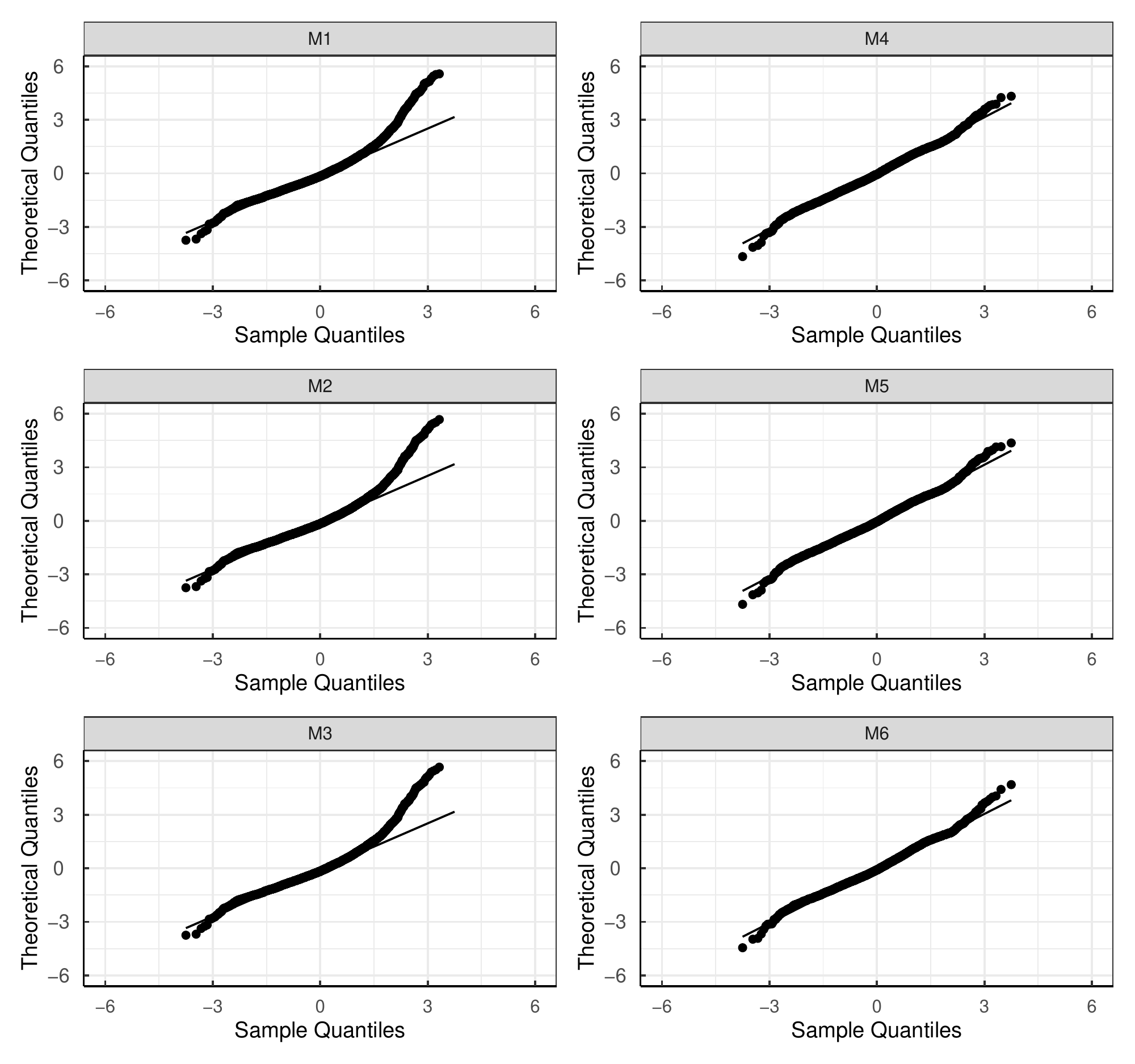}
	\caption{Medical claim dataset: QQ-plot for quantile residuals on the testing data.}
	\label{fig-medical-qq}
\end{figure}

\begin{figure}[htb]
	\centering
	\includegraphics[scale = 0.6]{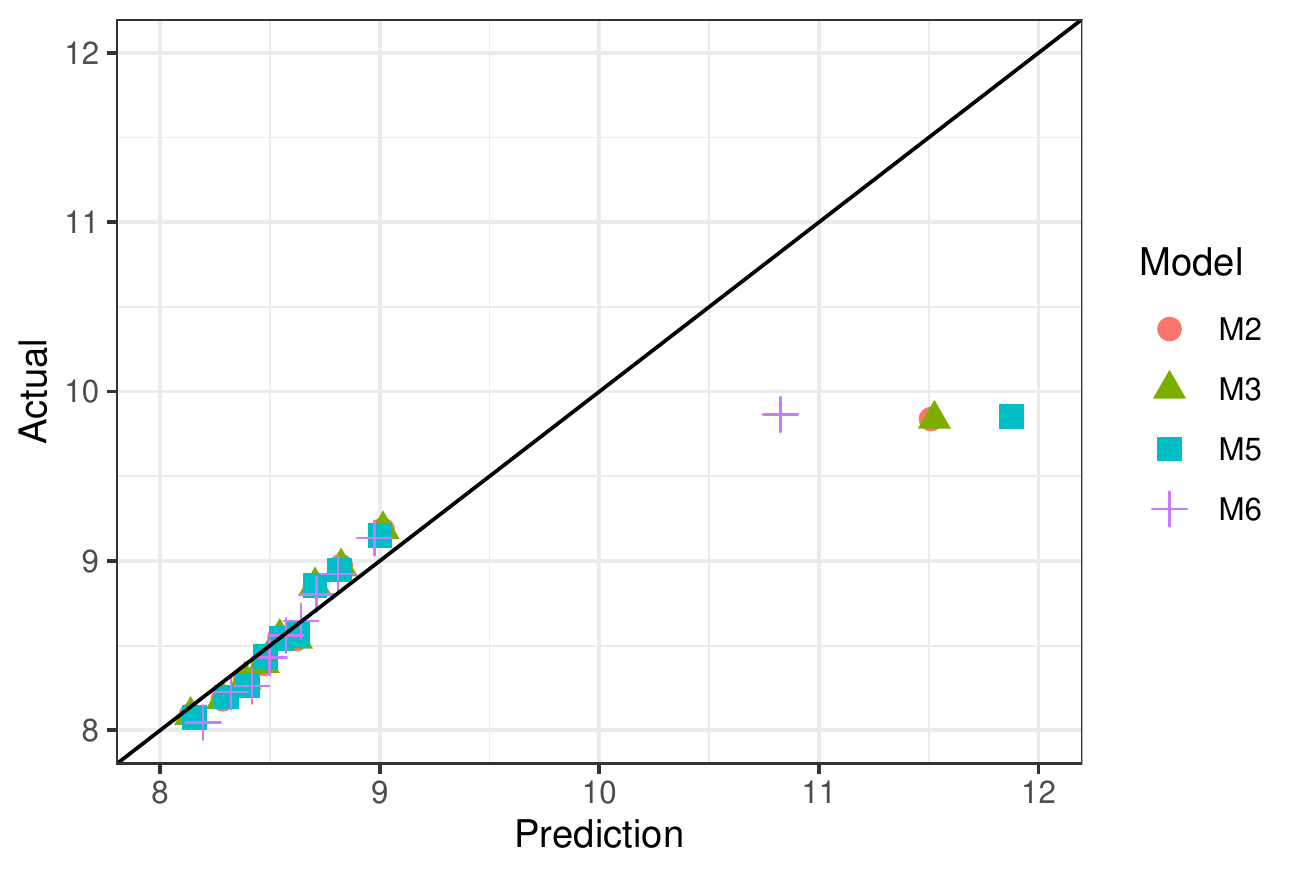}
	\caption{Medical claim dataset: the lift plot for GLMs without clustering method (M2),
		GLMs with clustering method (M3), mixture model without clustering method (M5) and
		mixture model with clustering method (M6) on the testing data.
	}
	\label{fig-medical-lift-plot}
\end{figure}

To illustrate the superiority of the proposed models, we consider fix competing regression models:
the GLMs without ICD (M1), the GLMs with ICD dummies (M2), the GLMs with ICD clustering (M3),
the mixture model without ICD (M4), the mixture model  with ICD dummies (M5), the mixture model with ICD clustering (M6).
Table \ref{table-medical-MSE} reports the NLL,
Pseudo $R^2$, MSE, MCRPS, and Lift metrics of the different fitted models on the testing data.
We select the optimal tuning parameter $v=0.1$ of M3 from $v\in \left\{0, 0.1, 0.2,\dots, 5\right\}$ based on the minimum NLL criteria on the training data. Our results show that the two-component mixture models (M4-M6) outperform the traditional GLMs (M1-M3) as evidenced by their lower NLL values. This suggests the presence of unobserved latent factors causing heterogeneity in claim amounts. The QQ-plots of the quantile residuals provided in Figure \ref{fig-medical-qq} support this finding, indicating a preference for the two-component mixture models in capturing the multi-modal and heavy-tailed features of medical claim amounts. Furthermore, the covariate clustering method for mixture models improves model performance in terms of out-of-sample MSE, with M6 having the lowest MSE. However, as expected, M6 does not perform well in terms of Pseudo $R^2$ and MCRPS metrics, while showing better ability to distinguish good risks from bad risks in terms of the lift metric. Figure \ref{fig-medical-lift-plot} illustrates this finding with lift plots for selected candidates. Overall, our results demonstrate the effectiveness of the proposed models in capturing the heterogeneity in medical claim amounts and improving model performance.

\begin{table}[htbp]
	\centering
	\caption{Medical claim dataset: out-of-sample performance for the six competing models.}
	\label{table-medical-MSE}
			\begin{tabular*}{\hsize}{@{}@{\extracolsep{\fill}}cccccc@{}}
		\toprule
		\multirow{2}[0]{*}{Models} & \multicolumn{5}{c}{Out-of-sample Performance} \\
		\cline{2-6}
		& NLL   & Pseudo R2 & MSE ($10^{11}$)   & CRPS & Lift \\
		\hline
        M1 & 54420.74  &\textbf{ 0.264}  & 1.15  & 0.0067427  & 5.89\\
M2 & {54426.70}  & 0.262  & 1.10  & 0.0067428  & 5.74\\
M3 & 54426.21  & 0.262  & 1.15  & \textbf{0.0067428}  & 5.75\\
\hline
M4 & 53935.07  & 0.260  & 2.37  & 0.0067474  & 5.83\\
M5 & 53945.57  & 0.259  & 2.88  & 0.0067475  & 5.91\\
M6 & \textbf{54007.50}  & {0.258}  &\textbf{ 0.18}  & {0.0067465}  & \textbf{6.16}\\
		\bottomrule
	\end{tabular*}
	\begin{tablenotes}
		\item *The optimal values of metrics are in bold.
	\end{tablenotes}
\end{table}

\clearpage
\section{Conclusion}\label{section-conclusion}

This paper discusses the framework for the implementation of covariate clustering in mixture regression models.
By adding a regularization term for similar covariates and using the EM-ADMM algorithm to solve the convex optimization problem, the proposed framework allows us to identify and analyze the clustering effects of covariates on heterogeneous subgroups of populations,
thus simultaneously exploring the complex network graph of the covariates and conducting feature selection.
The proposed method can also be considered as a convex clustering method for the simultaneous grouping of the covariate information and the outcome information.
Both simulation and empirical results show that our model can accurately identify the potential clusters of covariates and perform better than traditional (or mixture) GLMs in multiple datasets for the healthcare costs prediction.


Future research can explore the proposed framework from different angles. Firstly, while the Gamma distribution is utilized as the underlying distribution in the mixture model to model positive continuous response variables, it can be extended to other distributional assumptions. For instance, one may replace the Gamma distribution with Poisson or Tweedie distributions, to construct predictive models for the number of claims in the count type or the aggregate claim amounts containing zero values.
Secondly, while the two empirical examples demonstrated the application of the model to cluster covariates with similar observations and to cluster multiple levels of a factor-type variable, covariate clustering is also beneficial when reducing data dimensionality or characterizing the correlation relationship among covariates. Thus, the proposed model can be extended to more scenarios, such as spatio-temporal modeling based on various forms of penalties (e.g., well-known SCAD or MCP penalties).

Future studies can also explore how to enhance the interpretability of the proposed method, especially when covariate clustering is applied. It would be of interest to investigate whether the identified clusters can lead to meaningful insights into the characteristics of the population, or provide additional information for risk assessment. 
Furthermore, the proposed model can be further applied to other fields, such as finance, life insurance, and environmental science, where the heterogeneity of response variables is also commonly observed.


\section*{Acknowledgement}
Zhengxiao Li acknowledges the financial support from National Natural Science Fund of China (Grant No. 71901064) and ``the Fundamental Research Funds for the Central Universities" in UIBE (Grant No. CXTD13-02).
Yifan Huang acknowledges the financial support from the National Natural Science Foundation of China (Grant No. 72201062) and the National Social Science Fundation of China (Grant No. 22ATJ005).

\appendix
\section*{Appendices}

\addcontentsline{toc}{section}{Appendices}
\renewcommand{\thesubsection}{\Alph{subsection}}
\renewcommand{\theequation}{A.\arabic{equation}}
\subsection{The objective function in scaled ADMM}\label{section-appA}
At the M-step, we consider that the $w_h$ are constant in $Q(\bm{\Psi} ; \bm{\Psi}^{(m)})$ and maximize $Q(\bm{\Psi} ; \bm{\Psi}^{(m)})$ with respect to the other parameters in $\bm{\Psi}$ with the following form:
\begin{equation}
\label{eq-EM-M-optimization-0}
\mathrm{maximize}_{\beta_{0,h}, \bm{\beta}_h, \phi_h} \quad
\sum_{i=1}^n \pi_{i,h}^{(m)} \log{f\left(y_i | \mu_h(\bm{x}_i), \phi_h \right)}
- \omega \gamma {\left\| \bm{\beta}_h \right\|_{2}^{2}}
- \omega \frac{v}{2} \sum_{j,k \in I} s_{j,k} \abs{{\beta}_{j,h} - {\beta}_{k,h}},
\end{equation}
for $h=1,..,H$.
Note that in \eqref{eq-EM-M-optimization-0}, each set of parameters $\{ \beta_{0,h}, \bm{\beta}_h, \phi_h \}$ acts independently on the objective function, 
so the optimization problem can be decomposed according to the components and run in parallel. 

For simplicity we omit the subscript $h$ in the following because the updating process is consistent for $h=1,\dots,H$. 
The objective can be reformulated as a maximization problem in the following form:
\begin{equation}
\label{eq-EM-M-optimization}
\mathrm{maximize}_{\beta_{0}, \bm{\beta}, \phi} \quad
\sum_{i=1}^n \pi_{i}^{(m)} \log{f\left(y_i | \mu(\bm{x}_i), \phi \right)}
- \omega \gamma {\left\| \bm{\beta} \right\|_{2}^{2}}
- \omega \frac{v}{2} \sum_{j,k \in I} s_{j,k} \abs{{\beta}_{j} - {\beta}_{k}},
\end{equation}
summing up to obtain the overall maxima of $Q(\bm{\Psi} ; \bm{\Psi}^{(m)}) $.

Considering the ADMM algorithm  is well suited for decomposable dual optimization problems and has good convergence properties, we introduce a new set of auxiliary variables $\bm{z} = \{z_{j,k}\}$ in \eqref{eq-EM-M-optimization}.
The aim of M-step is to take the following form:
\begin{align}
\label{eq-ADMM-objective}
\mathrm{minimize}_{\beta_{0}, \bm{\beta}, \phi, \bm{z}} \quad
&- \sum_{i=1}^n \pi_{i}^{(m)} \log{f\left(y_i | \mu(\bm{x}_i), \phi \right)}
+ \omega \gamma {\left\| \bm{\beta} \right\|_{2}^{2}}
+ \omega \frac{v}{2} \sum_{j,k \in I} s_{j,k} \abs{z_{j,k} - z_{k,j}}, \nonumber\\
\text{s.t.} \quad & z_{j,k} - \beta_j = 0, \quad {\forall} j,k \in \{1,2,\dots,p\},
\end{align}
where $\bm{z} \in \mathbb{R}^{p\times p}$ constitutes a surrogate for the original regression coefficients $\bm{\beta} = (\beta_1, \beta_2, \dots, \beta_p)$.
The solution to this problem can be obtained by minimizing the following augmented Lagrangian function:
\begin{align}
\label{eq-ADMM-Lagrangian}
L_{\rho} \left(\beta_{0}, \bm{\beta}, \phi, \bm{z}, \bm{r} \right) =
&- \sum_{i=1}^n \pi_{i}^{(m)} \log{f\left(y_i | \mu(\bm{x}_i), \phi \right)}
+ \omega \gamma {\left\| \bm{\beta} \right\|_{2}^{2}} \nonumber\\
&+ \omega \frac{v}{2} \sum_{j=1}^p \sum_{k=1}^p s_{j,k} \abs{z_{j,k} - z_{k,j}} \nonumber\\
&+ \frac{\rho}{2} \sum_{j=1}^p \sum_{k=1}^p \left(z_{j,k} - \beta_j  + r_{j,k}\right)^2 + C,
\end{align}
where $\bm{r}=\left\{r_{j,k}\right\}\in \mathbb{R}^{p \times p}$ is a set of scaled dual variables,
$\rho>0$  is a positive penalty parameter and $C$ is a constant.
Note that \eqref{eq-ADMM-Lagrangian} is a scaled ADMM, see e.g. \cite{boyd2011distributed} in Section 3.1.1.
The ADMM algorithm is also widely discussed in recent studies \citep{chen2021identifying, zhang2022learning}.

\subsection{Update the regression coefficients}
In \eqref{eq-ADMM-beta},
when considering the commonly used log link function (i.e., $\mu_i = \mu(\bm{x}_i) = \exp(\beta_0 + \sum_{j=1}^p x_{i,j} \beta_{j})$), the derivatives of $g(\cdot)$ based on the chain rule are:
\begin{align}
\label{eq-ADMM-beta0-deriv}
\frac{\partial g(\beta_0, \bm{\beta}, \phi)} {\partial \beta_0}& =
-\sum_{i=1}^n \frac{f_{\mu}^{'}(y_i| \mu_i, \phi)} {f(y_i | \mu_i, \phi)} \mu_i,\\
\frac{\partial g(\beta_0, \bm{\beta}, \phi)} {\partial \beta_j} &=
-\sum_{i=1}^n \frac{f_{\mu}^{'}(y_i| \mu_i, \phi)} {f(y_i | \mu_i, \phi)} \cdot \mu_i \cdot x_{i,j}
+ 2 \omega \gamma \cdot \beta_j \nonumber \\
&- \rho \left[\sum_{k=1}^p \left(z_{j,k}^{(m,t)} + r_{j,k}^{(m,t)} \right.) - p \beta_j \right],
\quad {\forall} j = 1,2,\dots,p,
\label{eq-ADMM-betaj-deriv}
\end{align}
where $f_{\mu}^{'}(\cdot)$ represents the derivative function of the corresponding parametric distribution with respect to its mean parameter. The Broyden-Fletcher-Goldfarb-Shanno (BFGS) algorithm can be applied to calculate the updated regression coefficients $\beta_0^{(m,t+1)}, \bm{\beta}^{(m,t+1)}$ based on \eqref{eq-ADMM-beta0-deriv} and \eqref{eq-ADMM-betaj-deriv}, and $\phi^{(m,t+1)}$ can be updated by maximum likelihood estimation in numerical method.

\subsection{Update the auxiliary variables}\label{section-appD}
	
	According to \cite{hallac2015network}, this problem has a closed form solution; more specifically, for $j,k=1,2,\dots,p$, the updating formulas are:
	\begin{equation}
	\label{eq-ADMM-zjk-update}
	\left \{
	\begin{aligned}
	z_{j,k}^{(m,t+1)} = \theta_{j,k}^{(m,t+1)} \left(\beta_j^{(m,t+1)} - r_{j,k}^{(m,t)} \right) + (1-\theta_{j,k}^{(m,t+1)}) \left(\beta_k^{(m,t+1)} - r_{k,j}^{(m,t)} \right), \\
	z_{k,j}^{(m,t+1)} = (1-\theta_{j,k}^{(m,t+1)}) \left(\beta_j^{(m,t+1)} - r_{j,k}^{(m,t)} \right) + \theta_{j,k}^{(m,t+1)} \left(\beta_k^{(m,t+1)} - r_{k,j}^{(m,t)} \right),
	\end{aligned}
	\right.
	\end{equation}
	with
	\begin{equation}
	\label{eq-ADMM-zjk-theta}
	\theta_{j,k}^{(m,t+1)} =
	\begin{cases}
	& 0.5, \quad \text{if} \quad d_{j,k}^{(m,t+1)} \triangleq \sqrt{\left[(\beta_j^{(m,t+1)} - r_{j,k}^{(m,t)}) - (\beta_k^{(m,t+1)} - r_{k,j}^{(m,t)}) \right]^2} = 0, \\
	& \max{(1 - \frac{\omega v s_{j,k}}{\rho d_{j,k}^{(m,t+1)}}, 0.5)}, \quad \text{else}.
	\end{cases}
	\end{equation}
%

\subsection{The description of the questionnaire variables in diabetes expense dataset}\label{section-appC}
The following part is concerning some question statements for the questionnaire data, which are related to the knowledge of diabetes. 
In the V4 statement, it is posited that emotional stress can have an impact on blood sugar levels in individuals with diabetes. 
The V5 statement asserts that blood sugar levels may not necessarily increase the likelihood of developing complications associated with diabetes. 
The V77 statement suggests that smoking can significantly increase the risk of severe diabetic foot amputation in individuals with diabetes. 
Finally, the V79 statement notes that the combination of smoking and diabetes can lead to an increased risk of stroke. 
These statements highlight the complex interplay of factors that might affect the management and potential complications of diabetes, underscoring the importance of a holistic approach to treatment and risk reduction in individuals with this condition.

\clearpage
\bibliographystyle{plainnat}
\bibliography{mybibfile}
\end{document}